\documentclass[prb,twocolumn,showpacs,preprintnumbers,amsmath,amssymb,superscriptaddress]{revtex4-1}

\usepackage{graphicx}
\usepackage{dcolumn}
\usepackage{bm}

\def\Vec#1{\bm{#1}}

\begin{document}


\title{
Q-scan-analysis of the neutron scattering in iron-based superconductors
}

\author{Yuki Nagai}
\affiliation{CCSE, Japan  Atomic Energy Agency, 5-1-5 Kashiwanoha, Kashiwa, Chiba, 277-8587, Japan}
\affiliation{CREST(JST), 4-1-8 Honcho, Kawaguchi, Saitama, 332-0012, Japan}
\affiliation{TRIP(JST), Chiyoda, Tokyo 102-0075, Japan}
\author{Kazuhiko Kuroki}
\affiliation{Department of Applied Physics and Chemistry, The University of Electro-Communications, Chofu, Tokyo 182-8585, Japan}
\affiliation{TRIP(JST), Chiyoda, Tokyo 102-0075, Japan}


\date{\today}

\begin{abstract}
We propose a way to determine the pairing state of the iron pnictide superconductors exploiting the momentum ($\Vec{Q}$) scan of the neutron scattering data.
We investigate the spin susceptibility in the $s_{\pm}$ and $s_{++}$ 
superconducting states for various doping levels using the effective five-orbital model  and considering the quasiparticle damping. 
The peak position of the intensity shifts from the position on the line $Q_{x} = \pi$ to that on the line $Q_{y} = 0$ as the doping level is decreased from electron doping to hole doping. 
We find that the $\Vec{Q}$-dependence of the ratio of the intensity in the superconducting state to that in the normal state is qualitatively different between the $s_{\pm}$-wave and $s_{++}$-wave pairings.
We propose to investigate experimentally this ratio in Q-space to distinguish the two pairing states.
\end{abstract}

\maketitle

\section{Introduction}
The discovery of the iron-based superconductors 
has attracted considerable attention because of the 
high superconducting transition temperature\cite{Kamihara}. 
The possibility of a peculiar unconventional pairing state has also been an issue of great interest. 
In fact, spin-fluctuation-mediated $s_{\pm}$-wave pairing state has been proposed at the early stage of the
study \cite{BangPRB2008,SeoPRL2008,ParishPRB2008,KorshunovPRB2008,Mazin,KurokiPRL}, 
where the superconducting gap is fully open, but changes its sign across 
the wave vector that bridges the disconnected Fermi surfaces. 
The sign change occurs because repulsive pairing interaction arises from the spin fluctuations that develop around the Fermi surface nesting vector. 

There are many experimental results which suggest that the order parameter is fully gapped in a number of 
iron-based superconductors, such as the penetration depth measurements\cite{Hashimoto,LuetkensPRL2008,MaronePRB2009}, the angle resolved photo-emission spectroscopy\cite{DingEPL,Nakamura,Evtushinsky}, and the scanning tunneling microscopy/spectroscopy (STM/STS)\cite{HanaguriFe}. 
There are also experimental suggestions that the superconducting gap has
an unconventional form with sign change. 
The nuclear magnetic relaxation rate lacks the coherence peak below 
$T_{c}$\cite{Nakai,Grafe,MukudaJPSJ08,TerasakiFeNMR}, which 
suggests that there is a sign change in the order parameter. 
In addition, integer and half-integer flux-quantum transitions in composite niobium-ion pnictide loops have 
also been observed\cite{IBM}, which again suggests the presence of the sign change in the order parameter. 
In Ref.~\onlinecite{HanaguriFe}, STM/STS measurements have been performed to detect the quasiparticle interference originating from 
the $s_{\pm}$-gap. 
These experiments seem to be consistent with the $s_{\pm}$-wave pairing scenario. 

On the other hand, there has been a debate concerning the sensitivity of 
$T_{c}$ against impurities. 
There have been a number of studies regarding the issue of whether the pairing state in the iron pnictides is robust against impurities or not\cite{Guo,Nakajima,Kitagawa,Sato}. 
It has been pointed out that the suppression of $T_{c}$ by impurities is too week for a pairing state with 
a sign change in the gap in some papers e.g., Ref.~\onlinecite{Sato}. 
A calculation based on a five-band model by Onari {\it et al.} has supported this theoretically\cite{OnariKontani}, although the strength of the impurity potential adopted there is large compared to those 
calculated from first principles\cite{NakamuraArita}. 
As a possible pairing state that is robust agains impurities, the so-called $s_{++}$-state, 
where the gap does not change its sign between the Fermi surfaces, has been proposed\cite{Kontani,Ono,Saito}. 

As for the probe to determine the pairing state, it has been proposed at the early stage\cite{KorshunovPRB2008,Maier2,Maier} that 
the observation of neutron scattering resonance at the nesting vector of the electron and hole Fermi surfaces 
is one of the useful ways to determine whether there is sign change of the gap between these disconnected 
Fermi surfaces. 
In fact, neutron scattering experiments have observed a peak-like structure in the superconducting 
state\cite{Christianson,Qiu,Inosov,Zhao,Ishikado,Zhang}. 
This has been taken as strong evidence for the sign change in the superconducting gap. 
On the other hand, Onari {\it et al.} later took into account the quasiparticle damping effect in the calculation of the dynamical spin susceptibility, and 
showed that a peak-like enhancement over the normal state values can be obtained even in the $s_{++}$ state, 
which is due to the suppression of the normal state susceptibility originating from the damping\cite{Onari,Onari2}.  
In these studies, they claim that the enhancement that originates from the resonance of the $s_{\pm}$-wave pairing 
is too strong to explain the experimental observations, and the 'hump-like structure' of the intensity in the $s_{++}$-wave pairing is more likely to be the origin. However, a quantitative comparison between  theories and experiments should actually be very difficult since 
 there are many parameters that can affect the intensity, e.g., the doping-level, the band-structure, the strength of the electron interaction, and among all, the adopted theoretical method (random phase approximation, fluctuation exchange, etc.) Therefore, we need to look for a more qualitative difference between the
peak structures in the two pairing states, which can be understood intuitively.
Recently, we have proposed to investigate experimentally the wave vector $\sim (\pi,\pi)$ in the unfolded Brillouin zone, in addition to the usually considered $\sim(\pi,0)/(0,\pi)$,  to distinguish the two pairing states  qualitatively\cite{NagaiKuroki}. 
In this paper, we extend this study, and obtain a  $\Vec{Q}$
(momentum)-$E$(energy) scan of the dynamical susceptibility. The
momentum is scanned along the whole symmetric lines.
We also investigate the doping concentration dependence, where a recent
experimental as well as theoretical studies in the normal state have
revealed an electron-hole asymmetry of the incommensurability of the spin fluctuation\cite{LeePRL,Suzuki}. 
We propose that by looking at the superconducting to normal state ratio of the
dynamical susceptibility, the two states give qualitative difference 
reflecting the difference in the mechanism of the occurrence of 
the peak like structure.
The ratio is barely momentum dependent in the $s_{++}$ state, while 
in the $s_{\pm}$ state 
it is maximized around the wave vector at which the normal spin
susceptibility is maximized.

This paper is organized as follows. 
In Sec. II, four effective models for electron- or hole- doped systems and the
formulation of the spin susceptibility are presented. 
We introduce the multi-orbital random phase approximation (RPA) 
with the quasiparticle damping.
In Sec. III, the calculation results are presented.
We systematically investigate the doping-level and scattering vector dependences, and point out a qualitative difference between $s_{\pm}$-wave and $s_{++}$-wave states. In Sec. IV. the conclusion is given.

\section{Formulation}
\subsection{The effective models}

\begin{figure}[ht]
  \begin{center}
    \begin{tabular}{p{0.5 \columnwidth} p{0.5 \columnwidth}}
      \resizebox{0.5 \columnwidth}{!}{\includegraphics{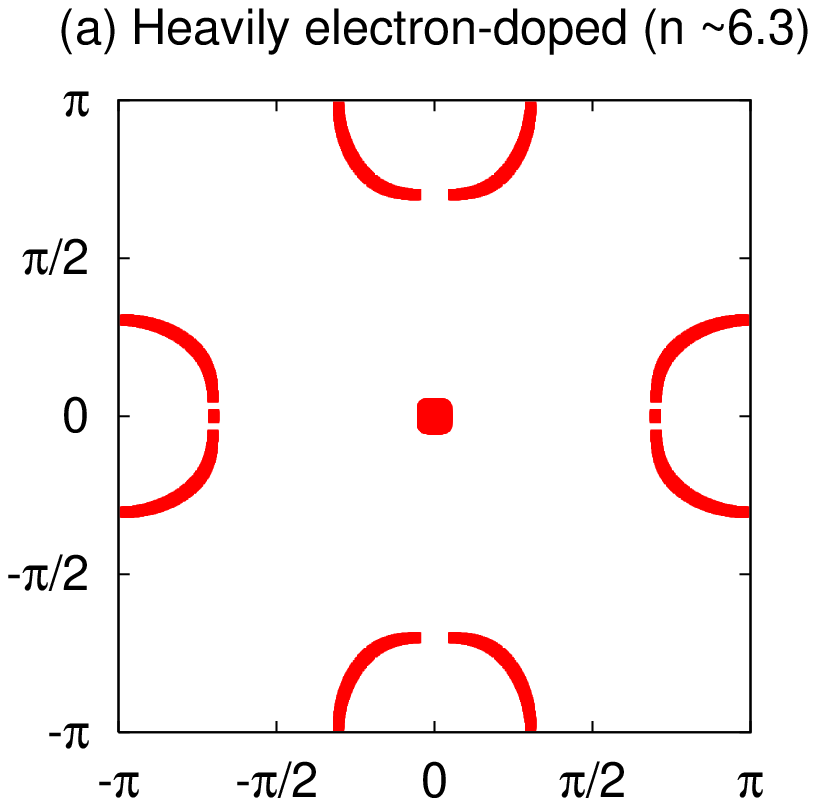}} &
      \resizebox{0.5 \columnwidth}{!}{\includegraphics{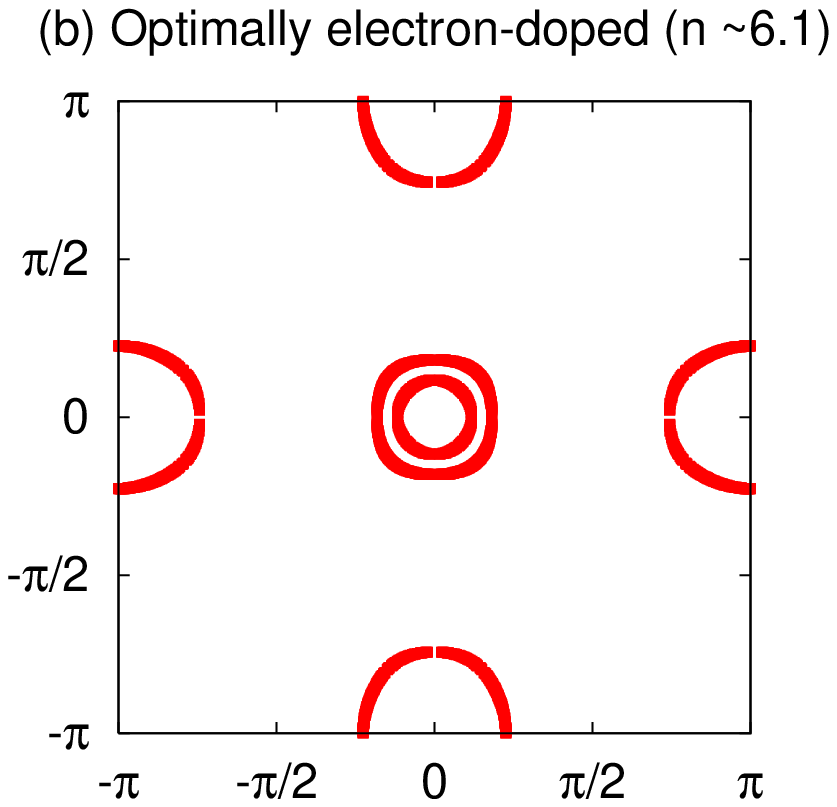}} 
    \end{tabular}
        \begin{tabular}{p{0.5 \columnwidth} p{0.5 \columnwidth}}
      \resizebox{0.5 \columnwidth}{!}{\includegraphics{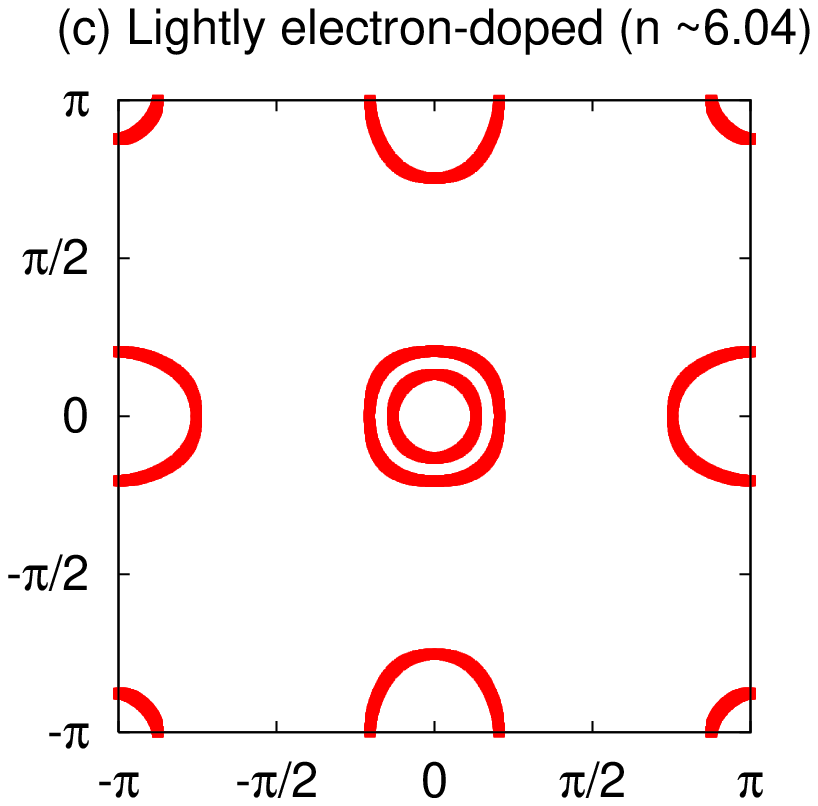}} &
      \resizebox{0.5 \columnwidth}{!}{\includegraphics{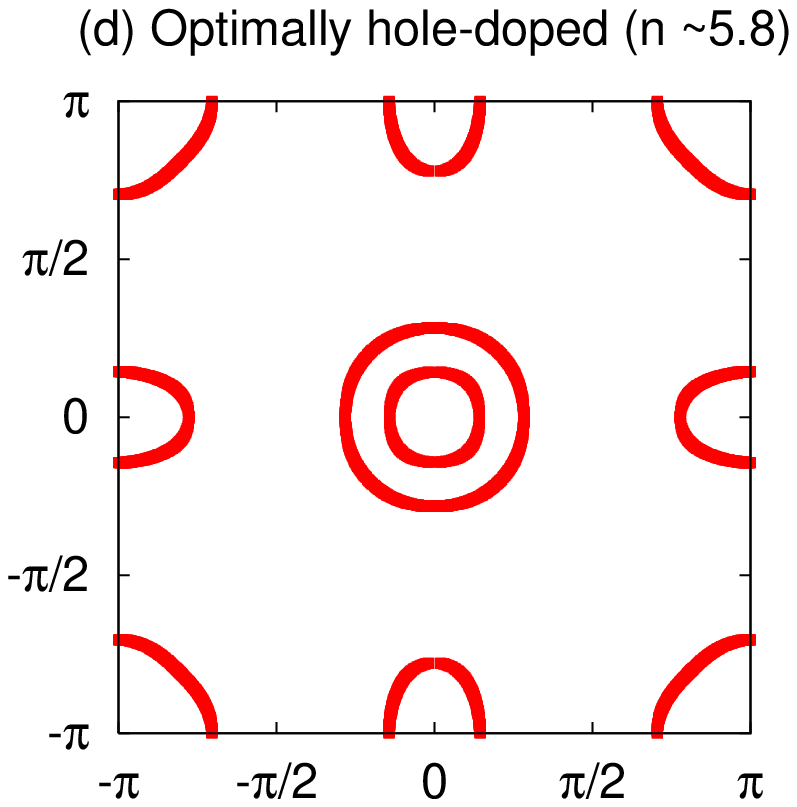}} 
    \end{tabular}
\caption{\label{fig:Fig1}
Fermi surfaces for (a) heavily electron-doped ($n \sim 6.3$), 
(b) optimally electron-doped ($n \sim 6.1$), 
(c) lightly electron-doped ($n \sim 6.04$),  
and (d) optimally hole-doped ($n \sim 5.8$) cases.
}
  \end{center}
\end{figure}

We introduce two-dimensional five-orbital models 
of 1111 materials 
obtained in the unfolded Brillouin zone\cite{KurokiPRL}, 
where the $x$- and $y$- axes are taken in the Fe-Fe bond direction. 
To investigate the doping-level dependence, 
we consider four situations : lightly electron-doped (the band filling $n \sim 6.04$) , 
optimally electron-doped ($n \sim 6.1$), heavily-electron doped (the band filling is $n\sim 6.3$), 
optimally hole-doped ($n \sim 5.8$) cases.
For the electron doped cases, we use the model of LaFeAsO, while for the 
hole doped case, to mimic the band structure of the 122 materials, 
where the $xy$ hole Fermi surface around the wave vector $(\pi,\pi)$ in the 
unfolded Brillouin zone is robust, we use a two dimensional 
model of NdFeAsO\cite{KurokiPRB}. 
This is just for simplicity in the
calculation, namely, the model of the 122 systems have 
strong three dimensionality, which is difficult to cope with within the 
present formalism.
We expect that the three dimensionality itself does not affect the 
conclusion of the present study.
The Fermi surfaces of each dopings are shown in Fig.~\ref{fig:Fig1}. 

\subsection{Spin susceptibility}
The dynamical spin susceptibility 
is given by 
\begin{align}
\chi(\Vec{Q}, E) = \sum_{a,c} \chi_{cc}^{aa}(\Vec{Q}, E). 
\end{align}
Here, $\chi^{aa}_{cc}(\Vec{Q},E)$ denotes an orbital-dependent spin susceptibility. 
The inelastic neutron scattering intensity is proportional to $\chi''(\Vec{Q},E)$ which is the imaginary part of $\chi(\Vec{Q}, E)$. 
In the multi-orbital RPA, an orbital-dependent spin susceptibility $\chi^{aa}_{cc}(\Vec{Q},E)$ 
is written as 
\begin{align}
\chi^{aa}_{cc}(\Vec{Q},E) &= \left[ (\hat{1} - \hat{\chi}_{0}(\Vec{Q},E)   \hat{U}_{s} )^{-1} \hat{\chi}_{0}(\Vec{Q},E)  \right]^{aa}_{cc},
\end{align}
where $\hat{\chi}_{0}(\Vec{Q},E)$ is the bare spin susceptibility expressed as
\begin{align}
\left[ \chi_{0}(\Vec{Q},E) \right]_{cd}^{ab} &= - \sum_{k} \sum_{\nu \nu'}  \left[
M_{abcd}^{\nu \nu' G}(\Vec{k},\Vec{k}+ \Vec{Q}) \chi_{0 G}^{\nu \mu}(\Vec{k},\Vec{k}+\Vec{Q},E) \right. \nonumber \\
& \left. + M_{abcd}^{\nu \nu' F} (\Vec{k},\Vec{k}+ \Vec{Q}) \chi_{0 F}^{\nu \nu'}(\Vec{k},\Vec{k}+\Vec{Q},E) \right] .
\end{align}
Here, $\chi_{0 G(F)}^{\nu \nu'}(\Vec{k},\Vec{k}+\Vec{Q},E)$ denotes the
normal (anomalous) part of the band-dependent BCS spin susceptibility 
written as 
%
%
%
\begin{align}
\chi_{0 G}^{\nu \nu'}(\Vec{k},\Vec{k}+\Vec{Q},E) &= \frac{|v_{\Vec{k}}^{\nu}|^{2}|u_{\Vec{k}+ \Vec{Q}}^{\nu'}|^{2}}{E + i \Gamma^{\nu \nu'}_{\Vec{k},\Vec{q}}- E_{\Vec{k} + \Vec{Q}}^{\nu'} - E_{\Vec{k}}^{\nu}  } \\
\chi_{0 F}^{\nu \nu'}(\Vec{k},\Vec{k}+\Vec{Q},E) &= - \frac{
u_{\Vec{k}}^{\nu \ast} v_{\Vec{k}}^{\nu} u_{\Vec{k}+\Vec{q}}^{\nu'} v_{\Vec{k}+\Vec{Q}}^{ \nu' \ast}
}{E + i \Gamma^{\nu \nu'}_{\Vec{k},\Vec{q}}- E_{\Vec{k} + \Vec{Q}}^{\nu'} - E_{\Vec{k}}^{\nu}  }, 
\end{align}
at zero-temperature ($E > 0 $) with 
$E_{\Vec{k}}^{\nu} = \sqrt{\epsilon_{k}^{\nu 2} + |\Delta_{k}^{\nu}|^{2}}$, 
$|u_{\Vec{k}}^{\nu}|^{2} = (1 + \epsilon_{\Vec{k}}^{\nu}/E_{\Vec{k}}^{\nu})/2$, 
$|v_{\Vec{k}}^{\nu}|^{2} = (1 - \epsilon_{\Vec{k}}^{\nu}/E_{\Vec{k}}^{\nu})/2$, 
$u_{\Vec{k}}^{\nu} v_{\Vec{k}}^{\nu \ast} = \Delta_{k}^{\nu}/(2 E_{\Vec{k}}^{\nu})$, and 
$\epsilon_{\Vec{k}}^{\nu}$ is the $\nu$-th band energy measured relative to the Fermi energy. 
$M_{abcd}^{\nu \nu' G}(\Vec{k},\Vec{k}+ \Vec{Q}) $ ($M_{abcd}^{\nu \nu' F}(\Vec{k},\Vec{k}+ \Vec{Q}) $) 
is given by 
\begin{align}
M_{abcd}^{\nu \nu' G}(\Vec{k},\Vec{k}+ \Vec{Q}) &= U_{a \nu}^{\ast}(\Vec{k}) U_{b \nu'}(\Vec{k} + \Vec{Q}) U_{c \nu'}^{\ast}(\Vec{k} + \Vec{Q}) U_{d \nu}(\Vec{k}), \\
M_{abcd}^{\nu \nu' F}(\Vec{k},\Vec{k}+ \Vec{Q}) &= U_{a \nu}^{\ast}(\Vec{k}) U_{b \nu'}(\Vec{k} + \Vec{Q}) U_{c \nu}(\Vec{k}) U_{d \nu'}^{\ast}(\Vec{k}+ \Vec{Q}), 
\end{align}
with the unitary matrix $\check{U}(\Vec{k})$ which diagonalizes the 
Hamiltonian in the orbital basis. 
Here, we introduce the band-index $\nu$ whose energy $\epsilon^{\nu}_{\Vec{k}}$ 
satisfies the relation $\epsilon^{\nu}_{\Vec{k}} > \epsilon^{\nu'}_{\Vec{k}}$ $(\nu > \nu')$.
For the '$s_{++}$-wave', we take  
$\Delta^{2} =  \Delta^{3 } = \Delta^{4}= \Delta_{0}$
 and for '$s_{\pm}$-wave' 
$\Delta^{2} =  \Delta^{3 } = -\Delta^{4}= \Delta_{0}$. 
As done in Ref.~\onlinecite{Maier}, we 
introduce a Gaussian cutoff for the gap 
 $\Delta_{\Vec{k}}^{\nu} = \Delta^{\nu} \exp \{-[\epsilon^{\nu}_{\Vec{k}}/\Delta E]^{2} \}$, 
and take $\Delta E = 4 \Delta_{0}$. 

We employ the orbital dependent interactions in the form 
$\hat{U}_{s} = a \hat{U}_{\rm Miyake}$. 
Here, $\hat{U}_{\rm Miyake}$ are the orbital dependent interactions obtained from the first-principles calculation by Miyake {\it et al}.\cite{Miyake}
In the RPA calculation, since realistic values of the interaction results in 
very large spin fluctuations,  we 
multiply all the electron-electron interaction by a reduction factor $a$. 
We set $a = 0.5$ for $n \sim 6.1$, $a = 0.45$ for $n \sim 6.04$, $a = 0.55$ for $n \sim 6.3$, and $a = 0.4$ for $n \sim 5.8$.
These values of $a$ are close to the maximum value that can be 
adopted in the present formalism (larger values give magnetic transition), 
which is dependent on the band filling. The variance of the maximum value of 
$a$ is actually largely due to the approximation adopted here (RPA), 
where the self energy corrections are 
not taken into account. Nevertheless,
we have confirmed that our results do not change qualitatively 
if we take other values of the  reduction factor.

We consider the quasiparticle damping $\Gamma^{\nu \nu'}_{\Vec{k},\Vec{q}}$ in 
the form 
\begin{align}
\Gamma^{\nu \nu'}_{\Vec{k},\Vec{q}} &= {\rm max} \: (\gamma_{\Vec{k}}^{\nu},\gamma_{\Vec{k}+\Vec{q}}^{\nu'}), 
\end{align}
where
\begin{align}
\gamma^{\nu}_{\Vec{k}} &= \left\{ \begin{array}{ll}
\eta & (|E_{\Vec{k}}^{\nu}| < 3 |\Delta_{k}^{\nu}| ) \\
\gamma_{0} & (4 |\Delta_{k}^{\nu}| < |E_{\Vec{k}}^{\nu}|) \\
(\gamma_{0} - \eta)\frac{|E_{\Vec{k}}^{\nu}|}{|\Delta_{k}^{\nu}|} - 3(\gamma_{0} -\eta) + \eta & ({\rm else}) \\
\end{array} \right. .
\end{align}
This form of the damping 
was introduced in Ref.~\onlinecite{Onari} from the requirement that 
the damping should not be present in the low energy regime of less than 
$3\Delta$, which is equal to the particle-hole excitation gap plus 
one-particle gap. In the high energy regime, the damping should be 
essentially the same as that in the normal state, and the present form 
interpolates the low and high energy regimes in a simple manner.
To be more strict, the damping of the superconducting state 
in the high energy regime  can be different from that in the 
normal state above $T_c$ because the temperature is different.
However, it is known experimentally that the dynamical susceptibility 
in the normal and the superconducting states nearly coincide with 
each other in the high energy regime\cite{Inosov}, indicating that 
the temperature dependence of the damping in the high energy 
regime is not so strong.
Therefore, we take the same value of $\gamma_0$ for the normal 
and the superconducting states.
Although we use this specific form of the damping in the actual 
calculation, the present results are not qualitatively dependent on the 
details of the damping.

\section{Results}
We investigate the $\Vec{Q}$-$E$ map of the spin susceptibility $\chi''(\Vec{Q},E)$. 
Since the commensurate vector is $Q = (\pi,0)$, 
we calculate $Q$-dependence along two lines in $Q$-map. 
One is the line on the $Q_{x}$-axis, i.e., $Q = (Q_{x},0)$, and 
the other is the line with $Q_{x} = \pi$,  $Q = (\pi,Q_{y})$.
In this paper, the smearing factor and the magnitude of the superconducting gap are taken to be $\eta = 0.5$ meV,  
$\Delta_{0} = 10$ meV, respectively. 
We take the quasiparticle damping in the normal state as 
$\gamma_{0} = 10$ meV, which was estimated from the temperature 
dependence of the resistivity observed experimentally\cite{Onari}.
To cope with the realistic magnitude of the superconducting gap, we take 
4096 $\times$ 4096 $\Vec{k}$-point meshes throughout the paper.

First, we will show the results in the normal state to see the peak
position of the spin susceptibility.
The incommensurability is found to be asymmetric between electron and
hole doped cases, as was found experimentally\cite{LeePRL} as well as theoretically\cite{Park2010,Suzuki}.
Then, we will present the results in the superconducting state to
investigate the difference between $s_{\pm}$-wave and $s_{++}$-wave states in $Q$-maps. 
We find that the $Q$-map of the superconducting/normal  ratio 
is qualitatively different between the two states.

\subsection{Normal state}
\begin{figure}[tb]
  \begin{center}
    \begin{tabular}{p{0.5 \columnwidth}p{0.5 \columnwidth}}
          \resizebox{0.48 \columnwidth}{!}{\includegraphics{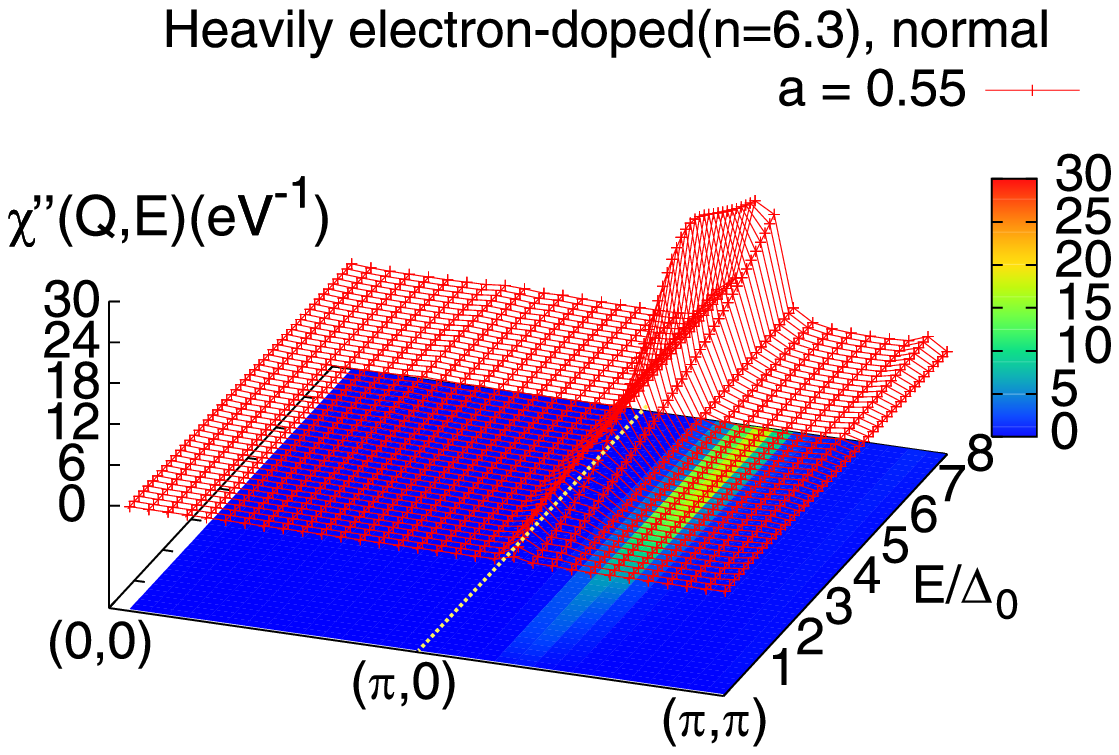}} &
      \resizebox{0.47 \columnwidth}{!}{\includegraphics{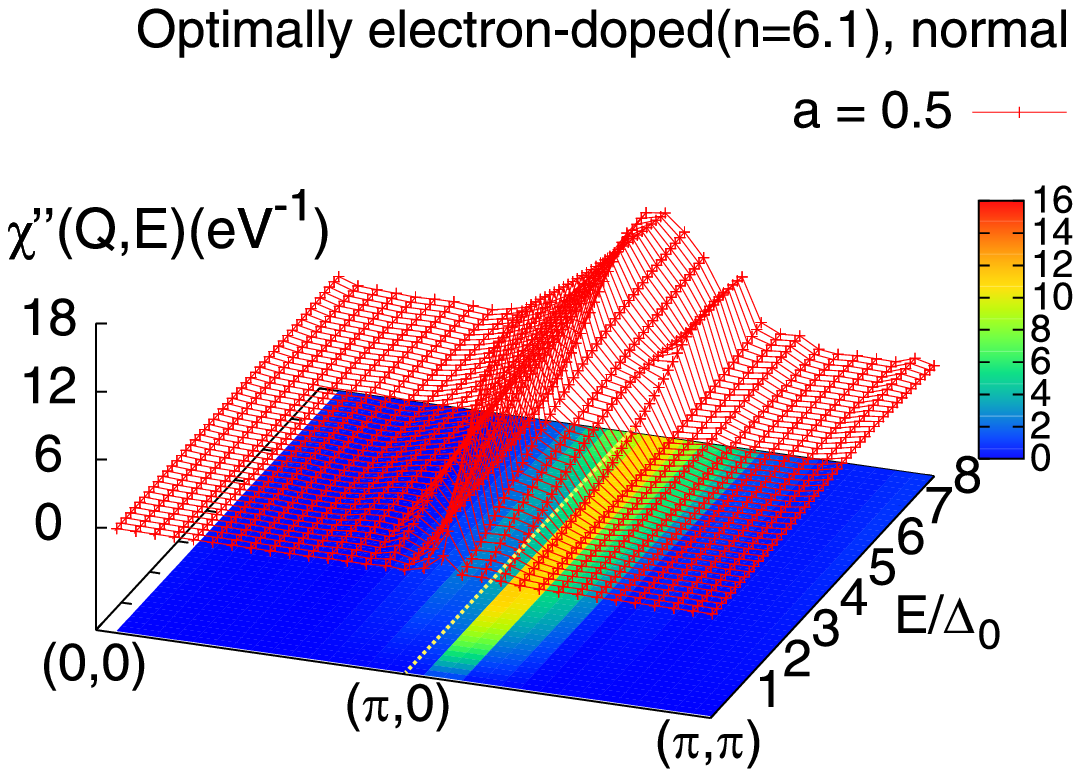}} \\
      \resizebox{0.48 \columnwidth}{!}{\includegraphics{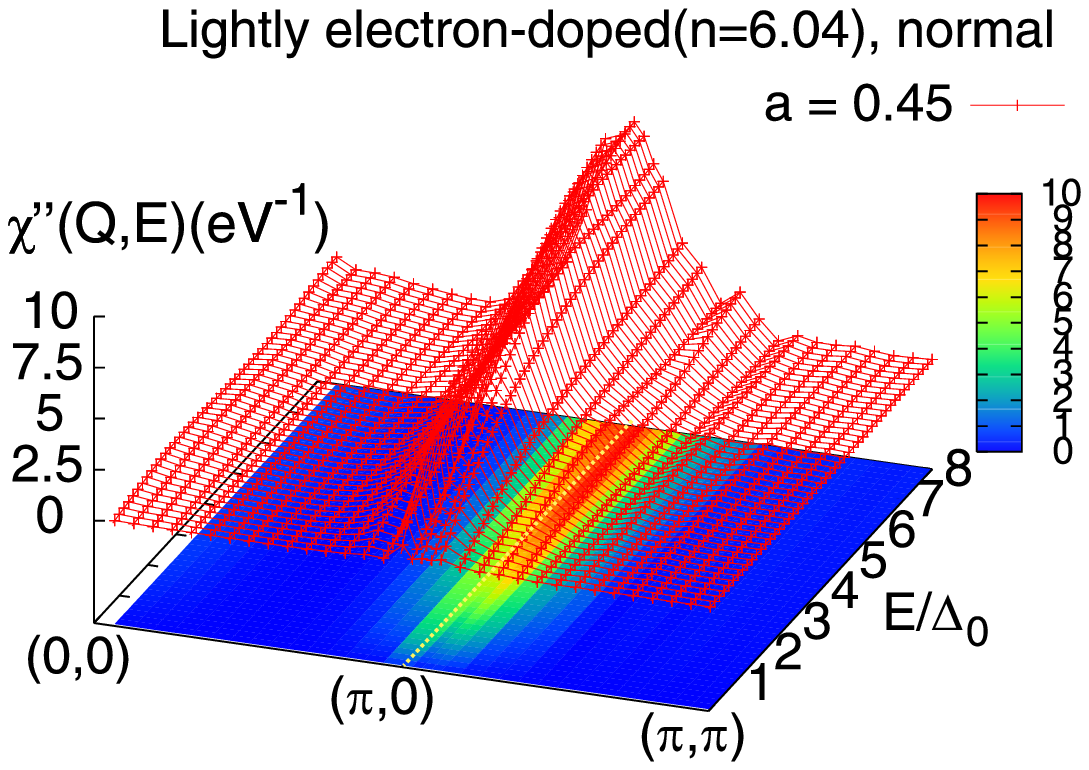}} &
      \resizebox{0.48 \columnwidth}{!}{\includegraphics{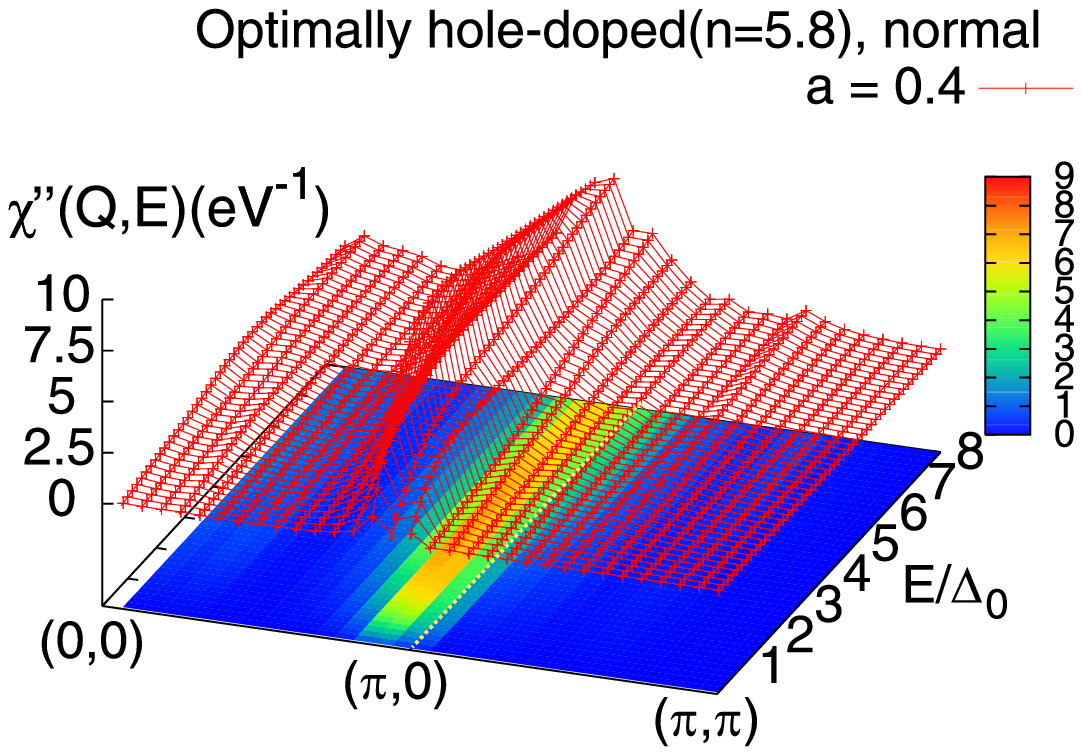}}
    \end{tabular}
\caption{\label{fig:Fig2}
$\Vec{Q}$-$E$ map of the spin susceptibility $\chi''(\Vec{Q},E)$ in the 
normal state for 
heavily electron-doped (the band filling is  $n \sim 6.3$), 
optimally electron-doped ($n \sim 6.1$), lightly electron-doped ($n \sim 6.04$), 
and optimally hole-doped ($n \sim 5.8$) cases. 
}
  \end{center}
\end{figure}
\begin{figure}[bt]
\begin{center}
\includegraphics[width = 0.5 \columnwidth]{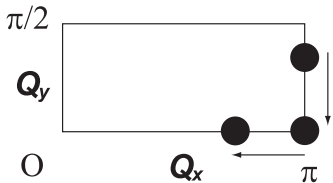}
\caption{\label{fig:Fig3}
Schematic figure of the peak-position shift in $Q$-space with decreasing the doping level.
}
\end{center}
\end{figure}

We show the results in the normal state in Fig.~\ref{fig:Fig2}. 
We find that the peak position of the intensity shifts from the position on the line $Q_{x} = \pi$ to that on the line 
$Q_{y} = 0$ as we decrease the electron doping level, and go into the
hole doping regime (see Fig.~\ref{fig:Fig3}). This electron-hole
asymmetry of the incommensurability of the spin fluctuations has been
attributed to the multiorbital nature of the Fermi
surface\cite{LeePRL,Suzuki}.
It should also be noted in the heavily electron doped regime of
$n=6.3$, the incommensurability is very large, coming somewhat close to $(\pi/2,\pi)$. 
This structure comes partially 
from the interaction between two electron pockets
\cite{KurokiPRL,KurokiPRB}, but there turns out to be also a contribution 
from the interaction between 
the electron and the barely present 
hole Fermi surfaces (as we shall see more clearly 
in the superconducting state), which are not well nested at 
this doping level. In this sense, we are using the term ``nesting'' in a
rather broad sense of the term. Another point to be noted in the 
heavily electron doped regime $(n=6.3)$ 
is the absence of intensity at low energies, 
which reflects the fact that the hole Fermi surfaces are barely present.
As for the interaction strength dependence, 
we have confirmed that the peak position does not depend on the 
interaction reducing factor $a$.


\subsection{Superconducting state: $s_{\pm}$-wave pairing}
\begin{figure}[t]
  \begin{center}
    \begin{tabular}{p{0.5 \columnwidth}p{0.5 \columnwidth}}
          \resizebox{0.48 \columnwidth}{!}{\includegraphics{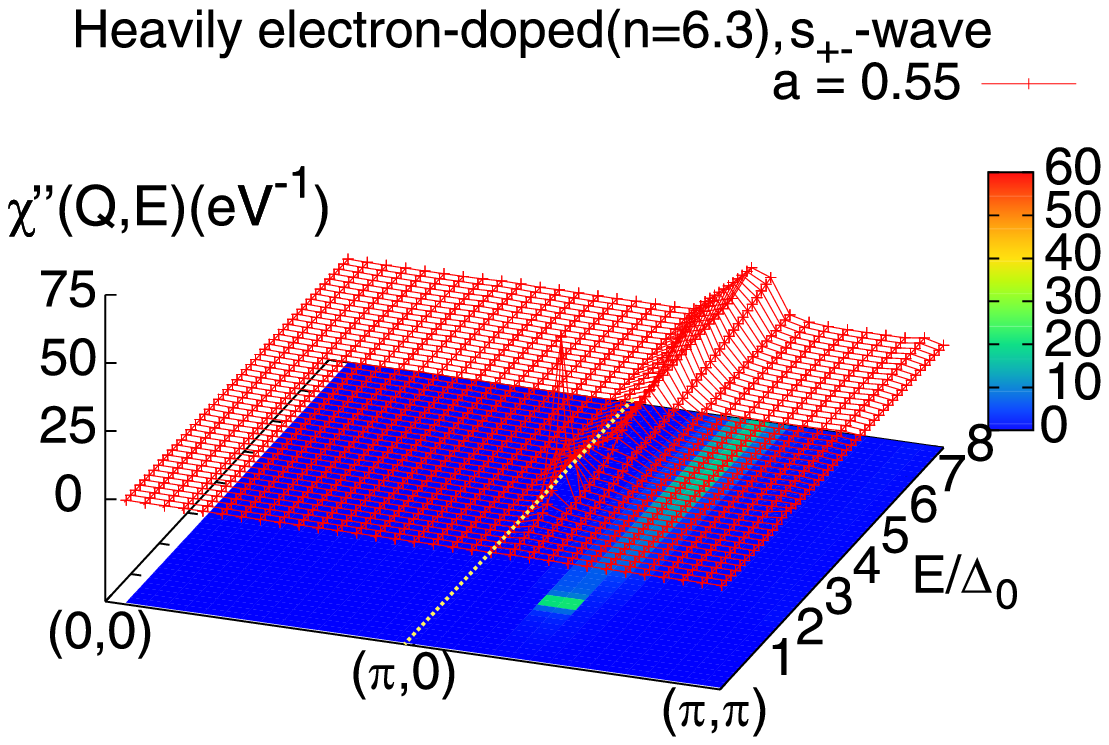}} &
      \resizebox{0.48 \columnwidth}{!}{\includegraphics{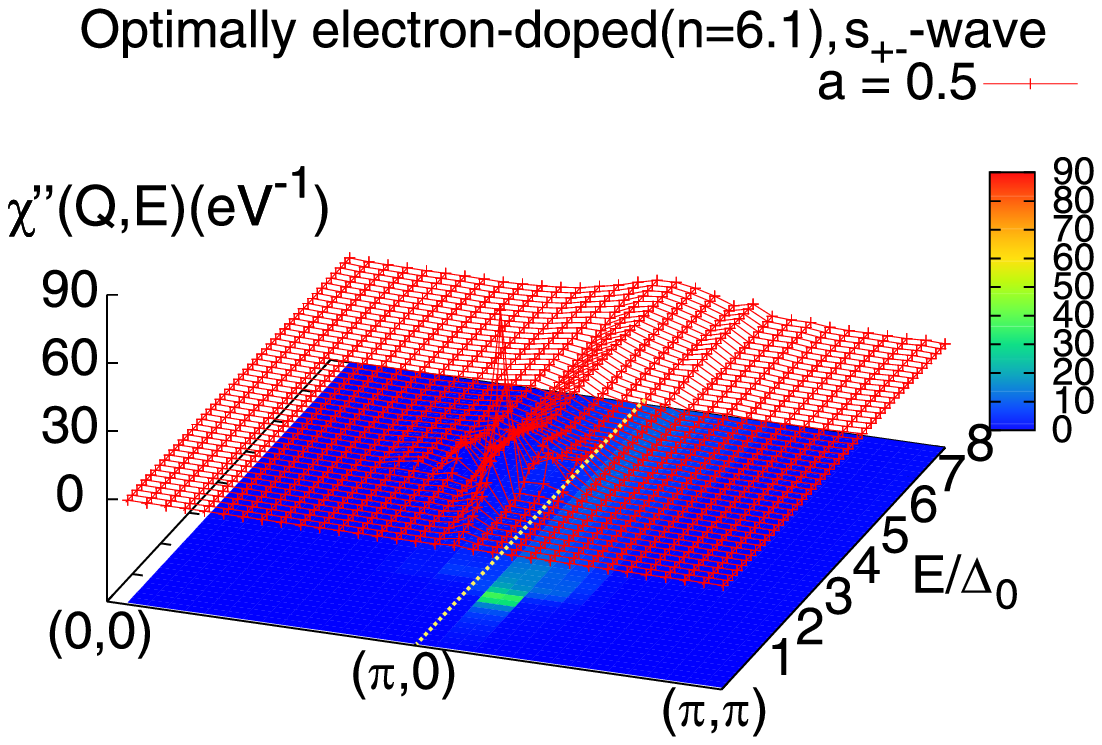}} \\
      \resizebox{0.48 \columnwidth}{!}{\includegraphics{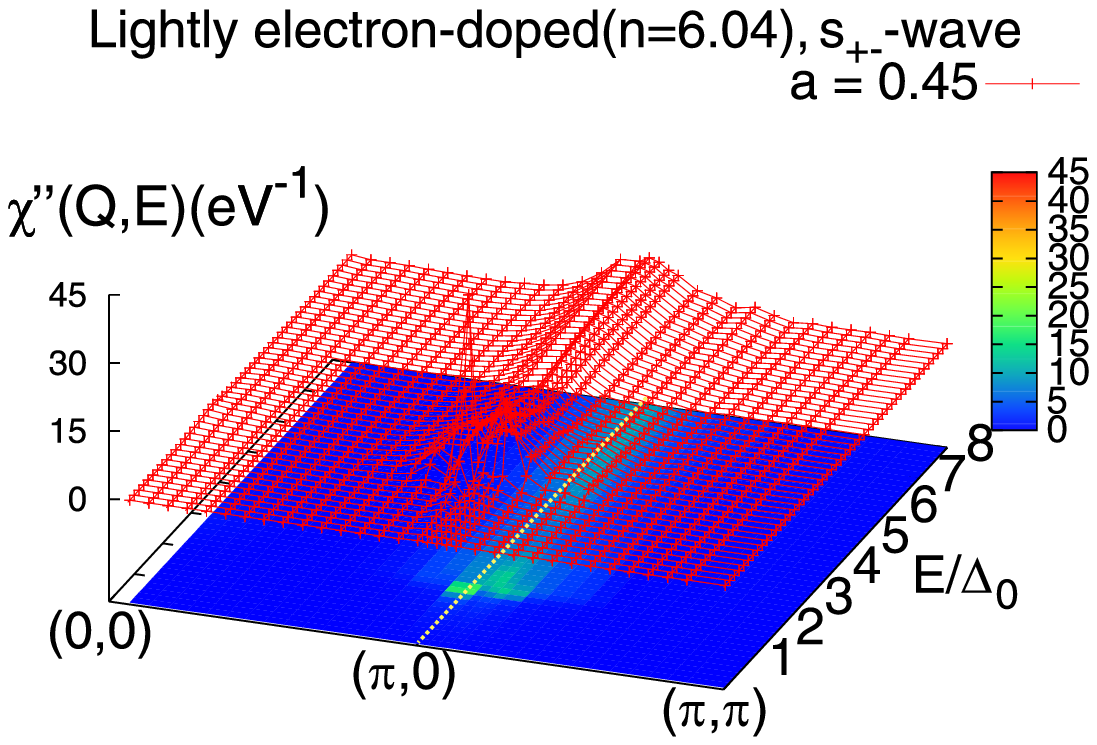}} &
      \resizebox{0.48 \columnwidth}{!}{\includegraphics{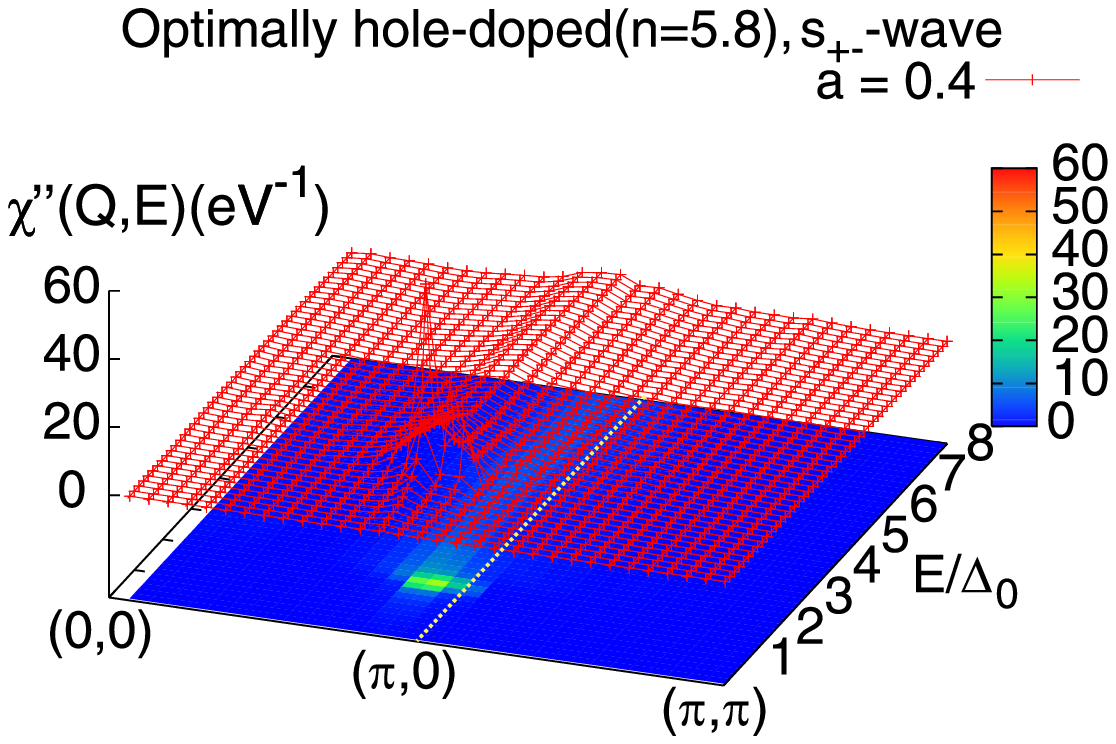}}
    \end{tabular}
\caption{\label{fig:Fig4}
$\Vec{Q}$-$E$ map of the spin susceptibility $\chi''(\Vec{Q},E)$ in the $s_{\pm}$-wave pairing state for the four band filling cases. 
}
  \end{center}
\end{figure}

Let us now consider the superconducting state with $s_{\pm}$-wave pairing. 
We show the $\Vec{Q}$-$E$ maps of the spin 
susceptibility $\chi''(\Vec{Q},E)$ in Fig.~\ref{fig:Fig4}. 
We can clearly see the shift of the peak position 
with decreasing the doping level, i.e., the resonance peak moves in 
accord with the normal state peak position.  
It should be noted that the peak intensity also 
depends on the doping level as well as the interaction strength. 
For example, the peak at the incommensurate vector $\Vec{Q} =
(\pi,\pi/8)$ for the optimally electron-doped case ($n \sim 6.1$) 
gives the largest peak intensity among these four doping levels. 
Therefore, we point out once again that it is difficult to determine 
the pairing state from quantitative comparison between experiments and theory 
since the peak intensity depends on the doping level $n$ 
as well as the strength of the electron interaction $\hat{U}_{s}$.  
Another point that should be noted here is that the resonance peak exists 
in the heavily electron doped regime $n=6.3$, 
reflecting the sign change of the gap 
between electron and hole Fermi surfaces. This means that there is a 
certain contribution to the structure close to $(\pi,\pi/2)$ also from the 
``badly nested'' electron and hole Fermi surfaces, as mentioned in the 
section for the normal state. A resonance at a wave vector 
close to  $(\pi,\pi/2)$ in heavily overdoped systems, as is observed 
for instance in Ref.\cite{ParkKeimer},  
is interpreted as a signature of $d$-wave pairing\cite{Maierdwave} where 
the gap sign changes between electron pockets, but the present study 
suggests that it may also be interpreted 
in terms of $s_{\pm}$ pairing originating from badly nested electron and 
hole pockets\cite{KAPRB}.

%
%
\begin{figure}[t]
  \begin{center}
    \begin{tabular}{p{0.5 \columnwidth}p{0.5 \columnwidth}}
          \resizebox{0.48 \columnwidth}{!}{\includegraphics{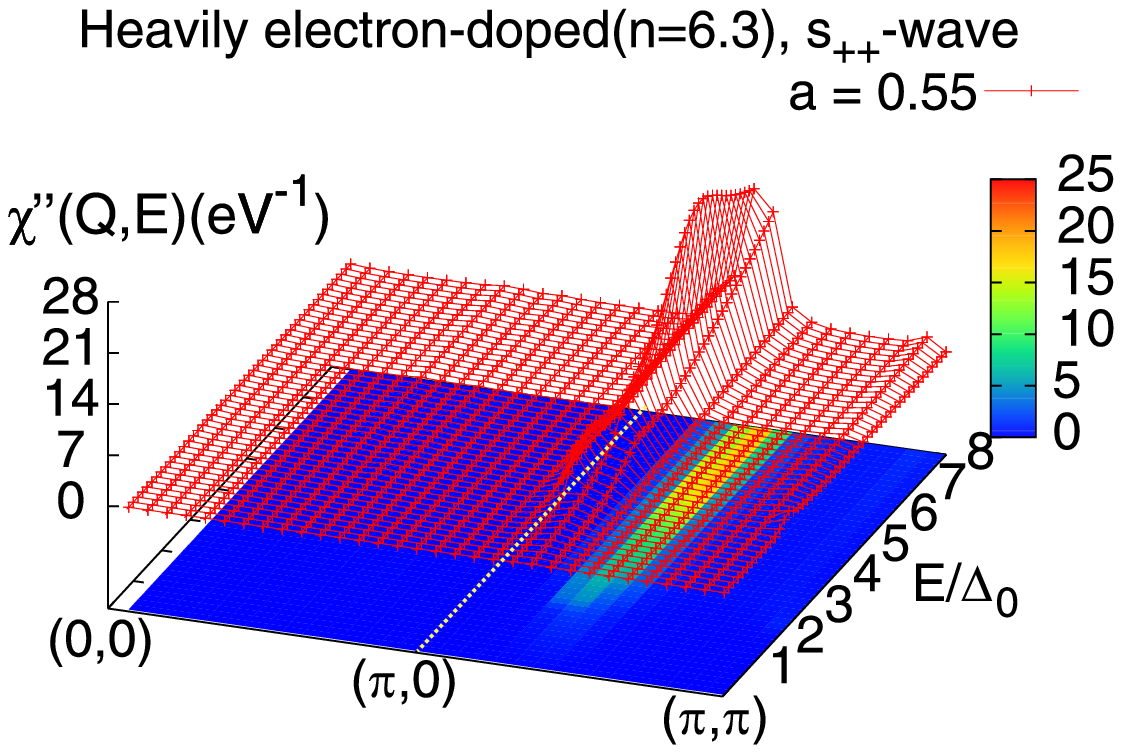}} &
      \resizebox{0.48 \columnwidth}{!}{\includegraphics{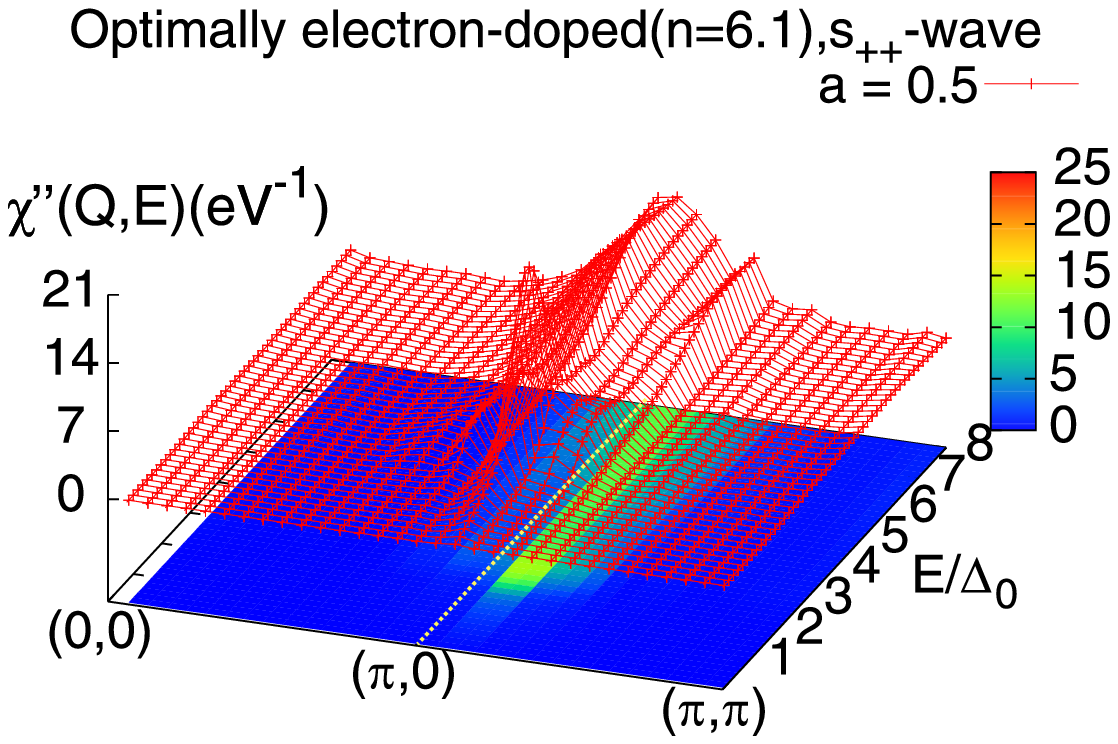}} \\
      \resizebox{0.48 \columnwidth}{!}{\includegraphics{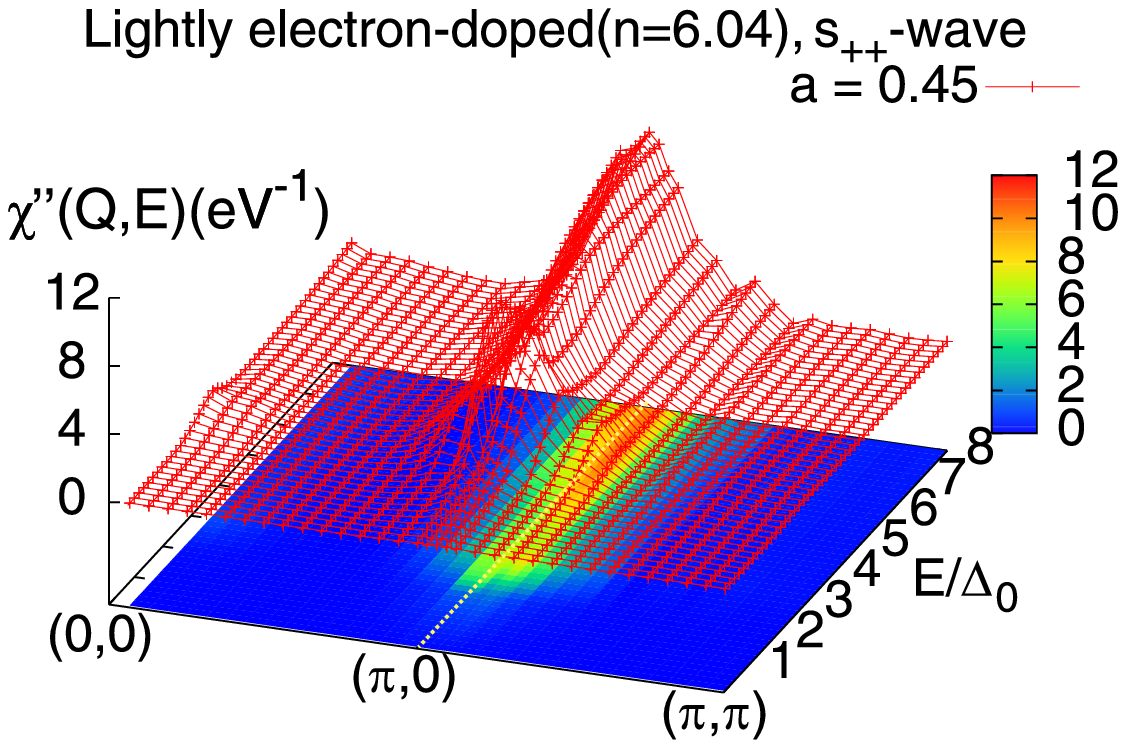}} &
      \resizebox{0.48 \columnwidth}{!}{\includegraphics{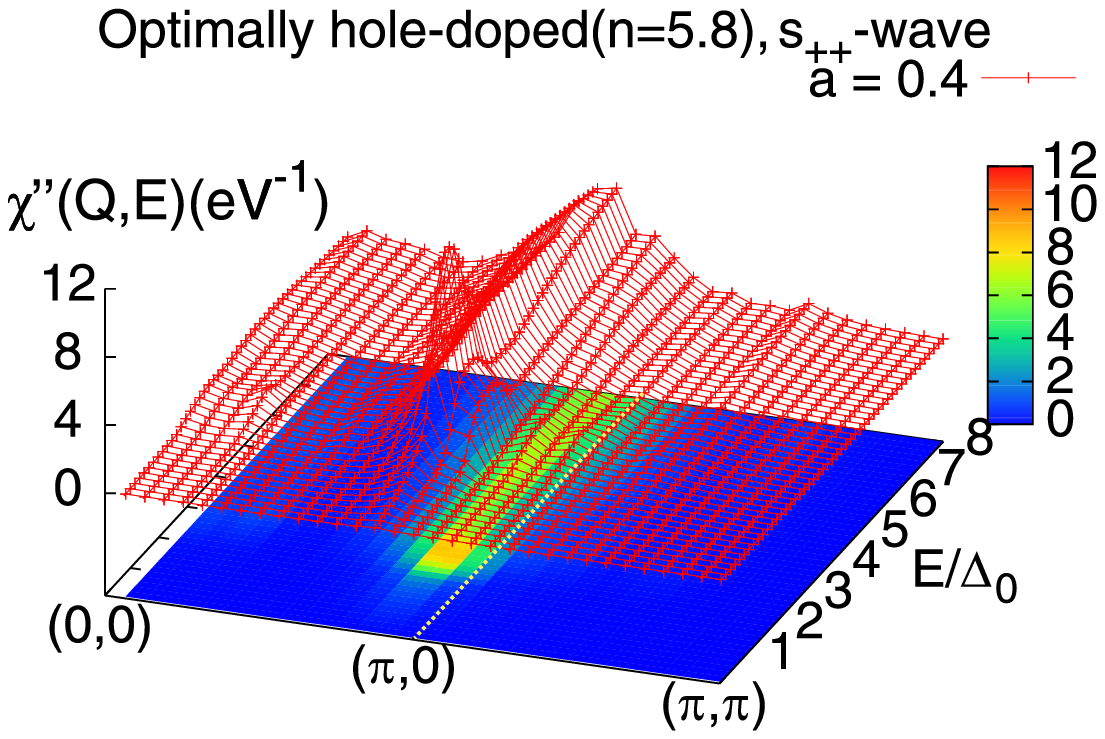}}
    \end{tabular}
\caption{\label{fig:Fig5}
$\Vec{Q}$-$E$ map of the spin susceptibility $\chi''(\Vec{Q},E)$ in the $s_{++}$-wave pairing state for the four band filling cases.
}
  \end{center}
\end{figure}
\subsection{Superconducting state: $s_{++}$-wave pairing}
Next, we consider the $s_{++}$-wave state. 
We show the $\Vec{Q}$-$E$ maps of the spin susceptibility $\chi''(\Vec{Q},E)$ in 
Figs.~\ref{fig:Fig5}. 
As pointed out previously\cite{Onari,NagaiKuroki}, 
the quasiparticle damping makes the hump structure in the $s_{++}$-wave pairing state. 
The hump structure moves as the doping level is varied, which 
at first sight may look similar to that in the $s_{\pm}$ state. 
However, there is a large difference between these two states, which is 
revealed by taking the ratio between the normal and the superconducting states.

\begin{figure}[t]
  \begin{center}
    \begin{tabular}{p{0.5 \columnwidth}p{0.5 \columnwidth}}
      \resizebox{0.48 \columnwidth}{!}{\includegraphics{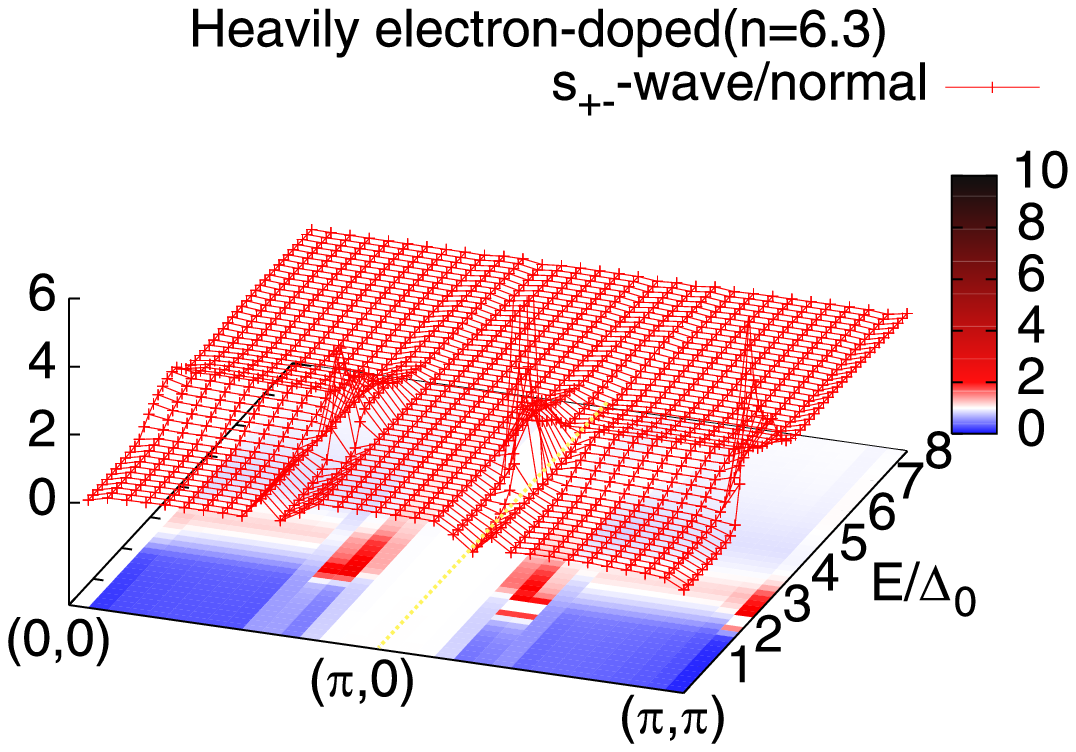}} &
      \resizebox{0.48 \columnwidth}{!}{\includegraphics{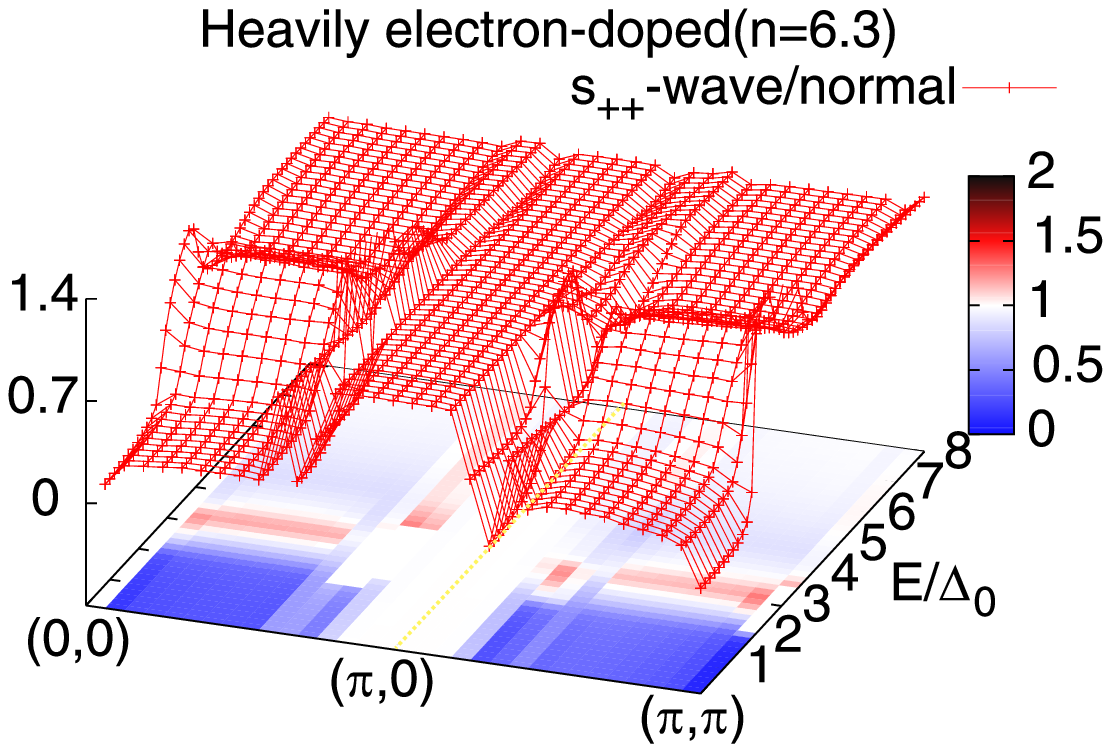}}   \\
            \resizebox{0.48 \columnwidth}{!}{\includegraphics{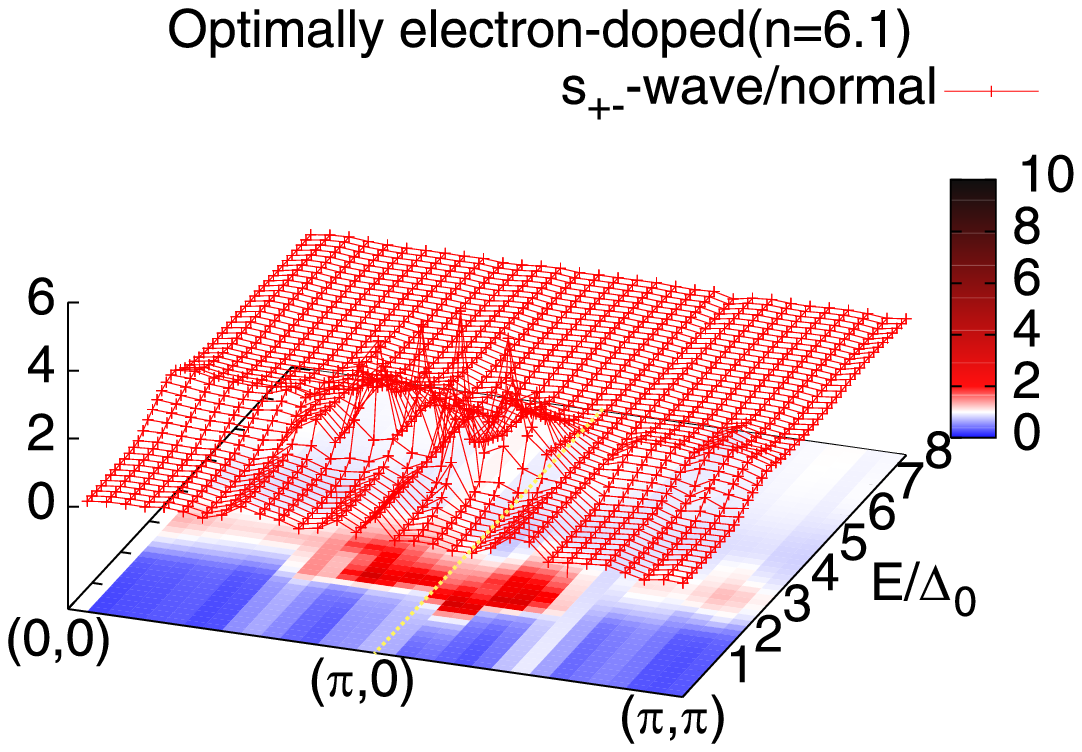}} &
      \resizebox{0.48 \columnwidth}{!}{\includegraphics{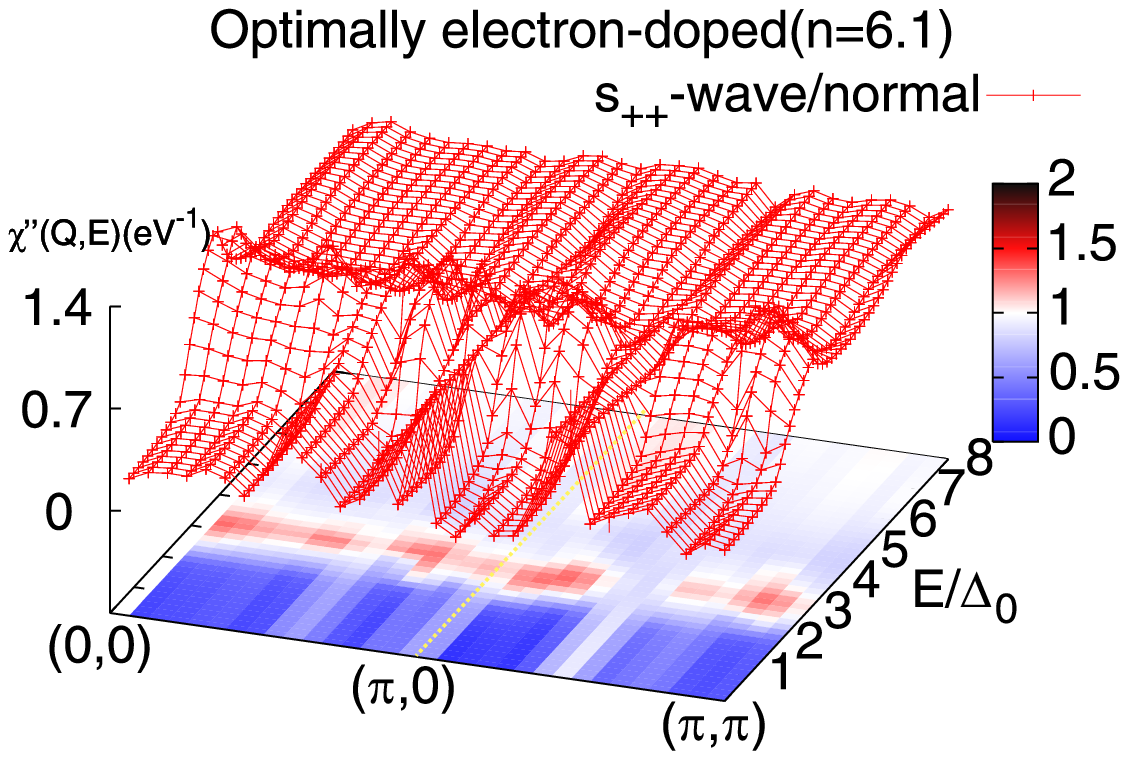}} \\
            \resizebox{0.48 \columnwidth}{!}{\includegraphics{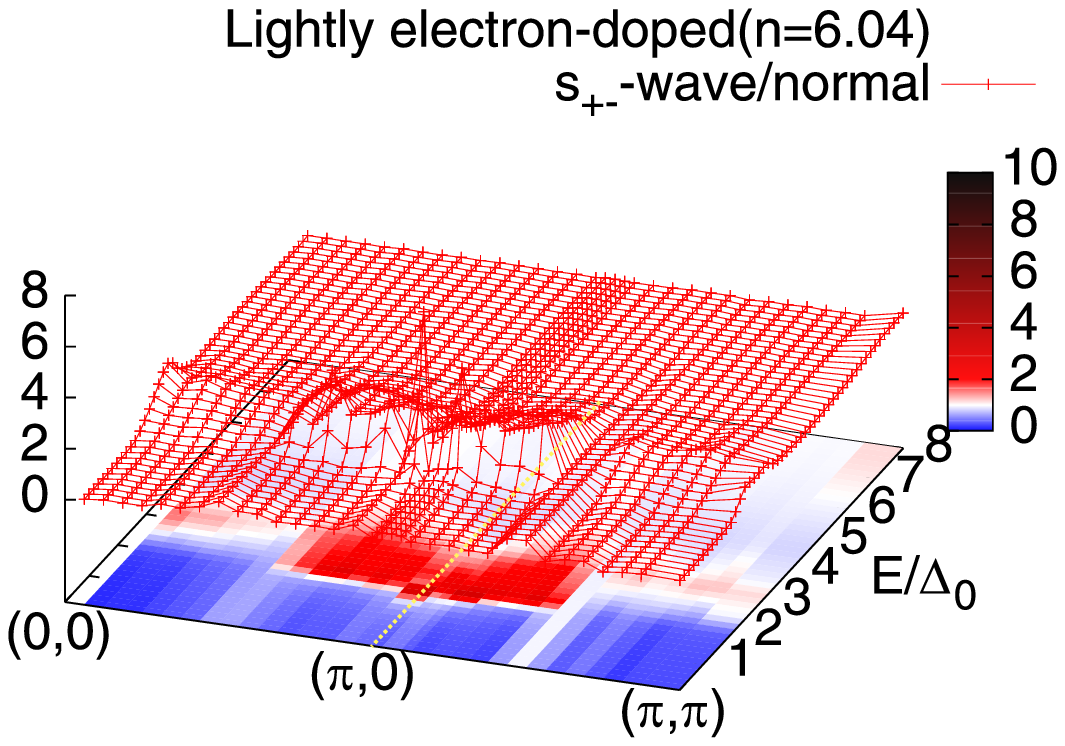}} &
      \resizebox{0.48 \columnwidth}{!}{\includegraphics{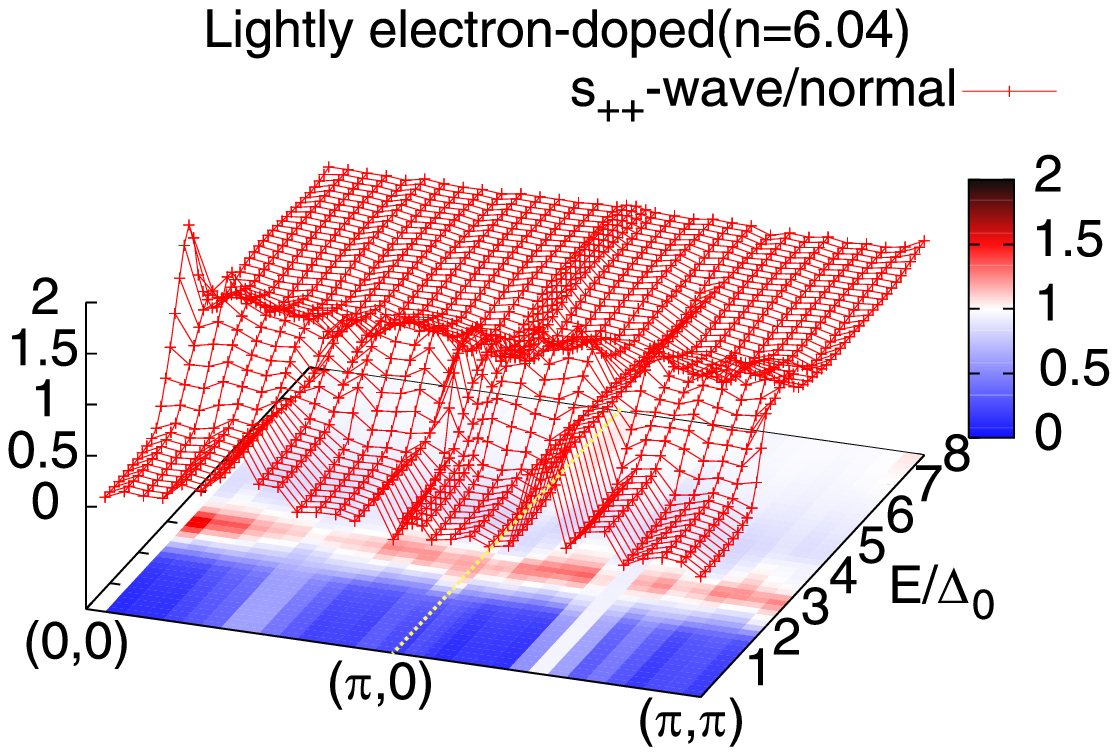}} \\
      \resizebox{0.48 \columnwidth}{!}{\includegraphics{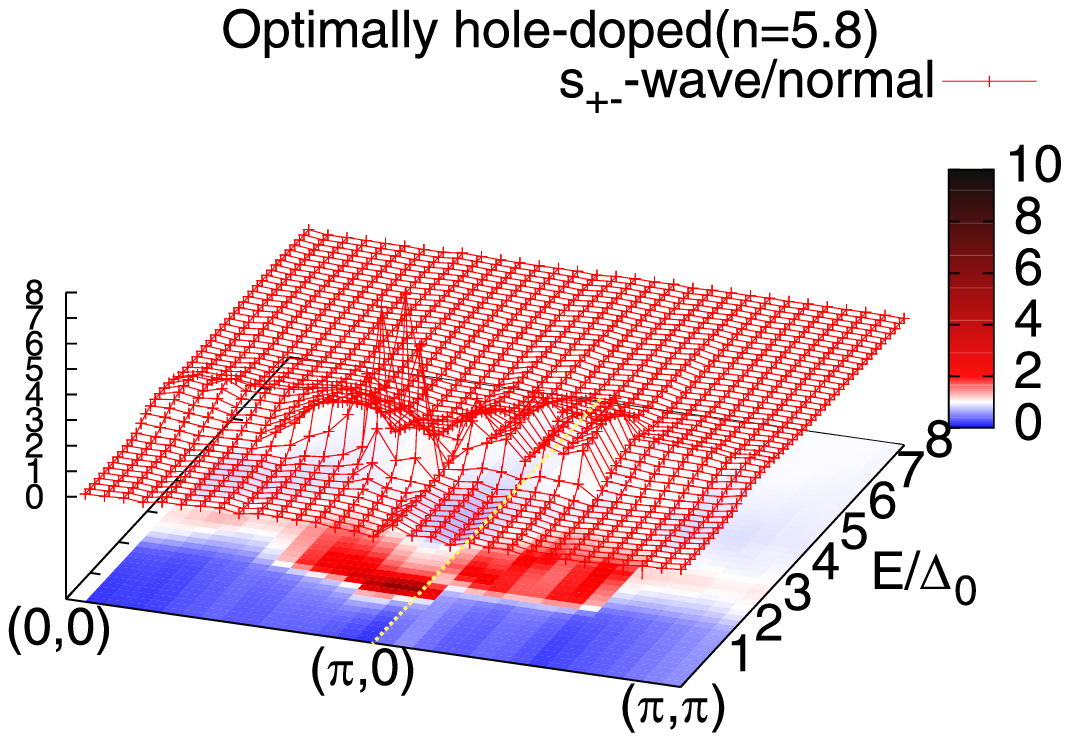}} &
      \resizebox{0.48 \columnwidth}{!}{\includegraphics{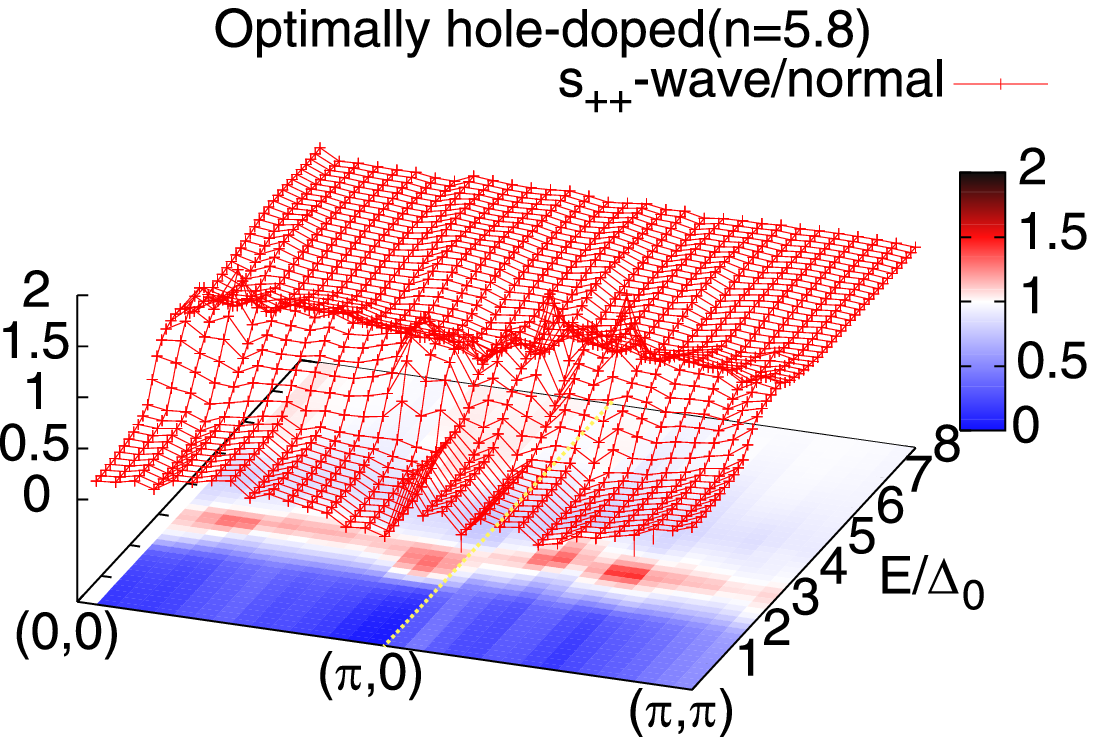}} 
    \end{tabular}
\caption{\label{fig:Fig6}
$\Vec{Q}$-$E$ map of the ratio of the intensity in the superconducting 
state to that in the normal state for the four band filling cases.
Left panel: the result of the $s_{\pm}$-wave pairing state. 
Right panel: the result of the $s_{++}$-wave pairing state.
}
  \end{center}
\end{figure}
\subsection{Superconducting to normal ratio}
\begin{figure}[bt]
  \begin{center}
    \begin{tabular}{p{1 \columnwidth}}
      \resizebox{0.5 \columnwidth}{!}{\includegraphics{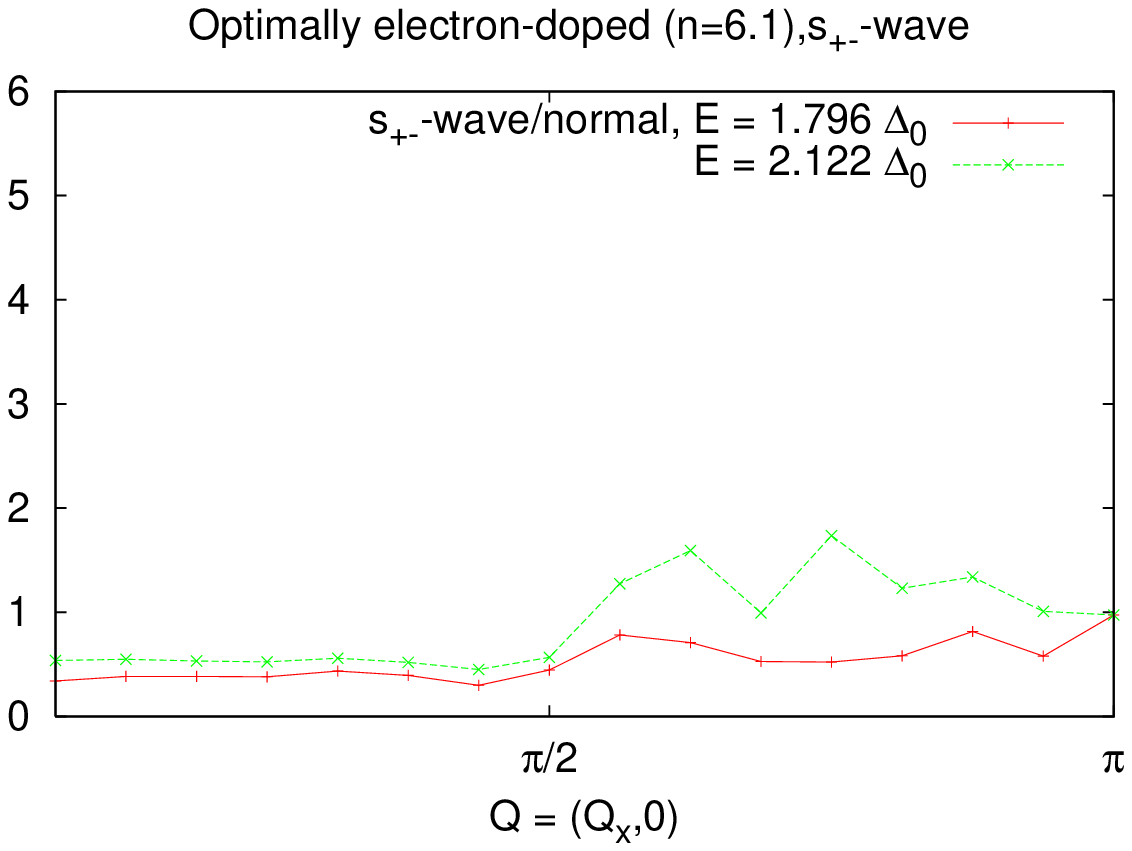}} 
      \resizebox{0.5 \columnwidth}{!}{\includegraphics{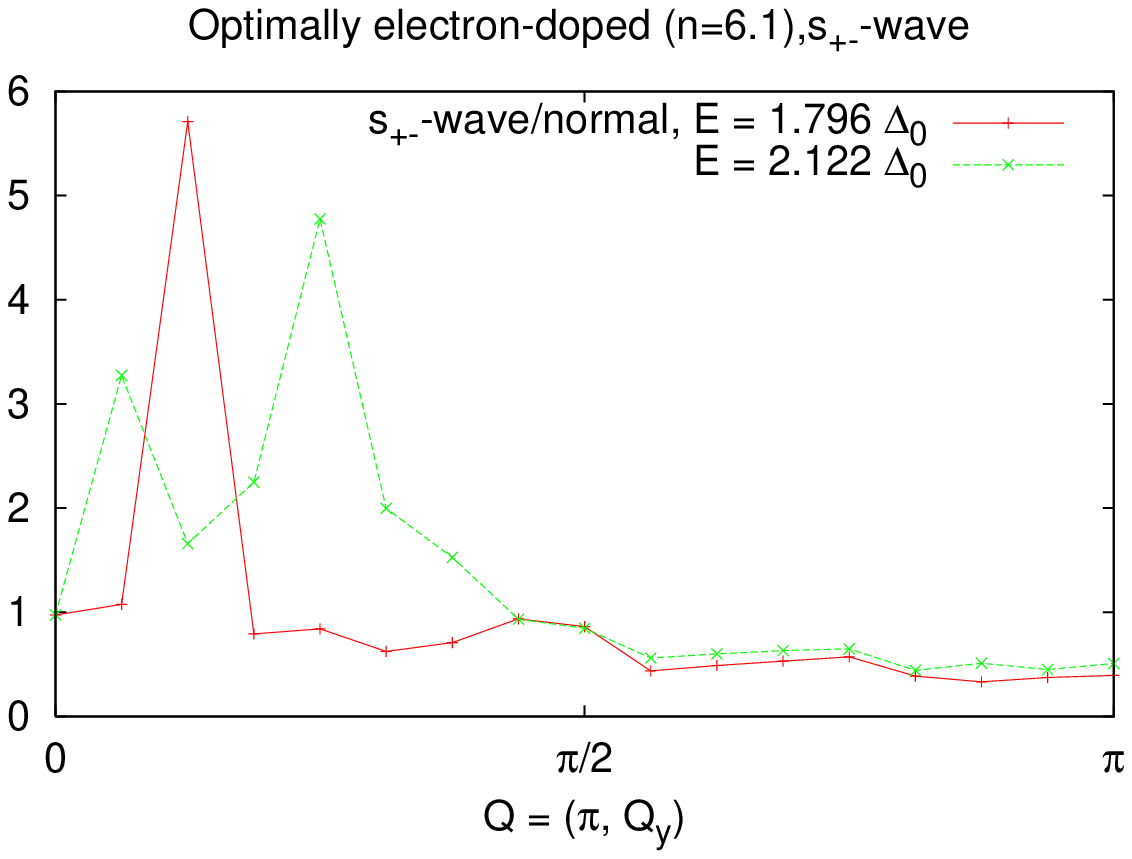}} \\
            \resizebox{0.5 \columnwidth}{!}{\includegraphics{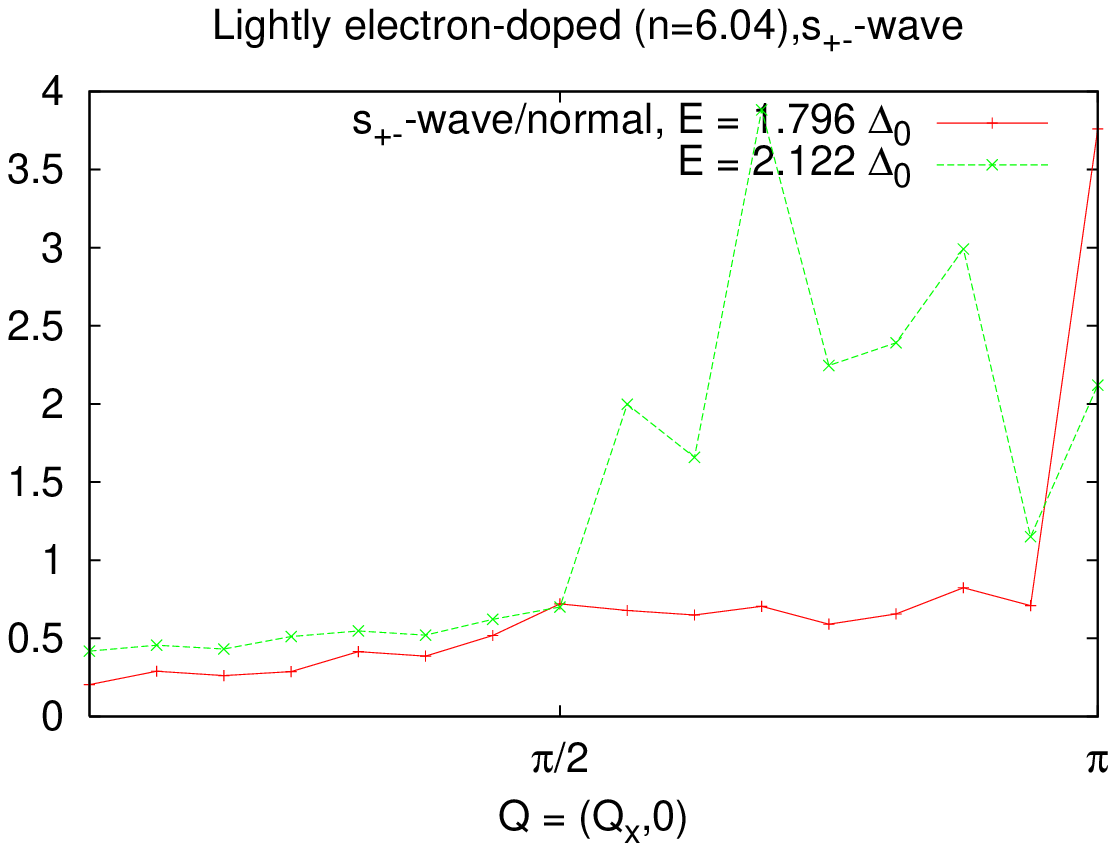}} 
      \resizebox{0.5 \columnwidth}{!}{\includegraphics{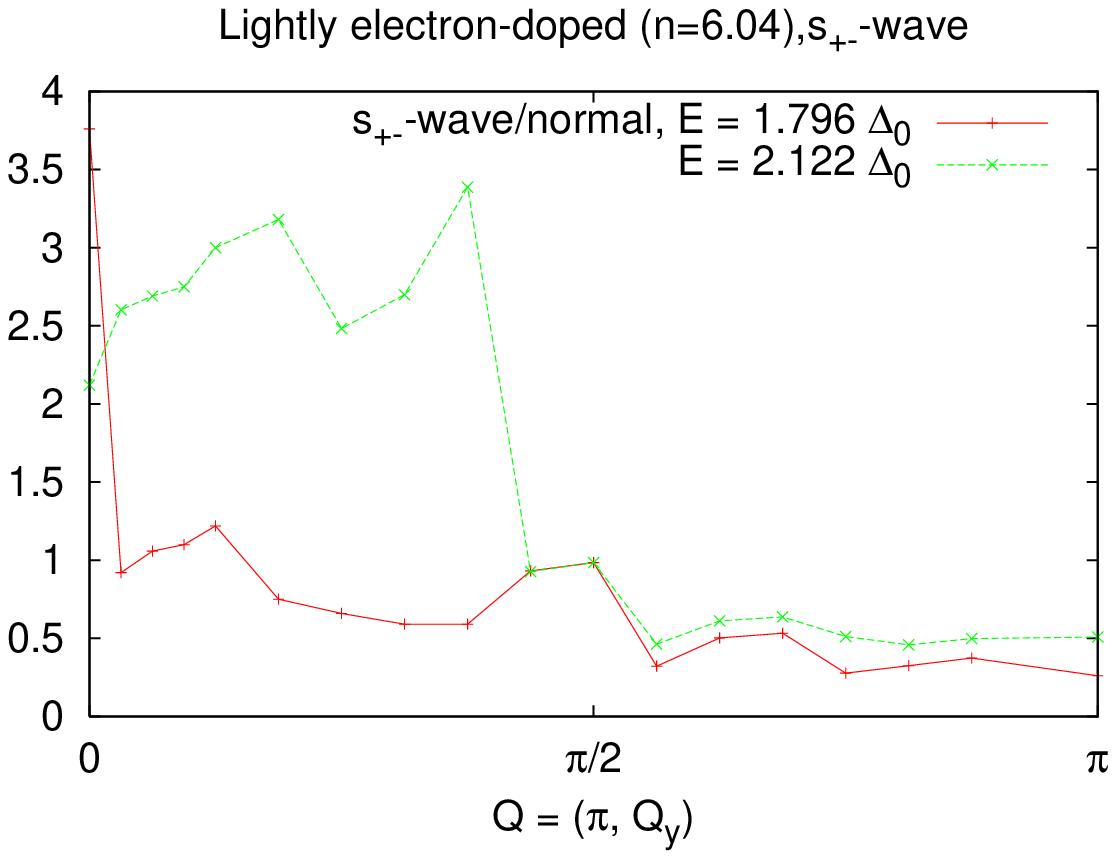}} \\
            \resizebox{0.5 \columnwidth}{!}{\includegraphics{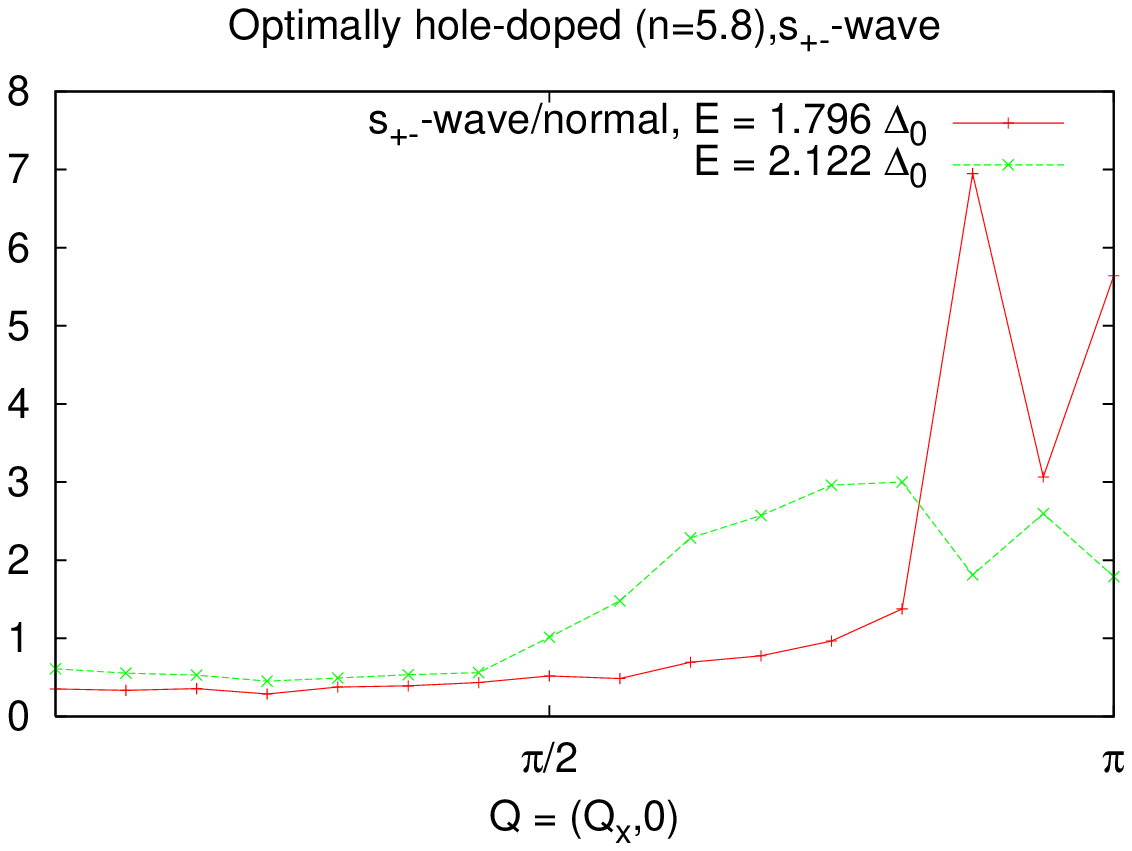}} 
      \resizebox{0.5 \columnwidth}{!}{\includegraphics{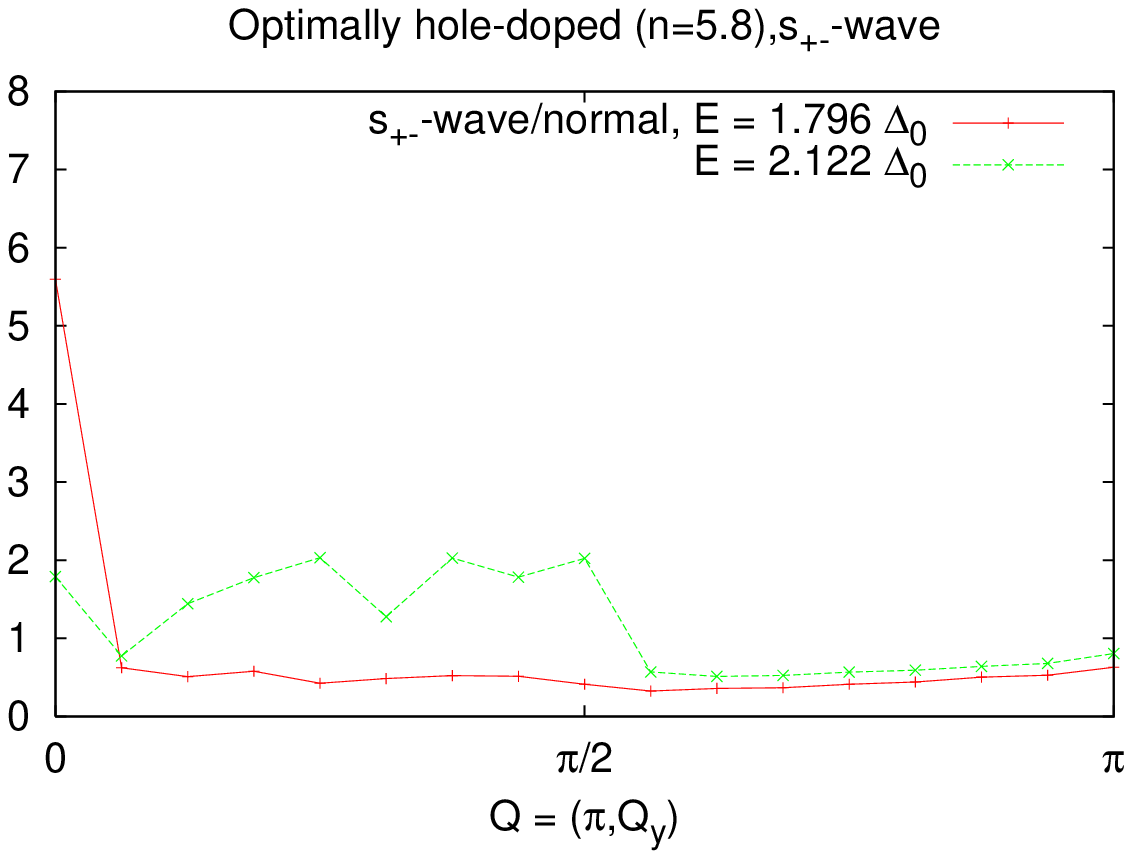}} 
    \end{tabular}
\caption{\label{fig:Fig10}
$\Vec{Q}$-dependence of the $\chi''(\Vec{Q},E)$ in the $s_{\pm}$-wave pairing state. 
Left panel: $Q_{x}$-dependence $\Vec{Q} = (Q_{x},0)$. 
Right panel: $Q_{y}$-dependence $\Vec{Q} = (\pi,Q_{y})$.
}
  \end{center}
\end{figure}

We now 
show the $\Vec{Q}$-$E$ map of the ratio of the spin susceptibility in
the superconducting state to that in the normal state in Figs.~\ref{fig:Fig6}. 
As shown in the left panels in these figures, the ratio becomes large
in a localized regime around  $\Vec{Q} \sim (\pi,0)$ in the case of the $s_{\pm}$-wave pairing. 
To be more precise, 
one can find in Fig.~\ref{fig:Fig10} that the peak position of the ratio 
in the low energy region ($E < 2 \Delta_{0}$) shifts 
with decreasing the doping level, in accord with the peak position of
the spin susceptibility in the normal state as was shown in Fig.~\ref{fig:Fig3}. 
By contrast, in the case of $s_{++}$-wave pairing, we find that the
ratio is weakly dependent on $\Vec{Q}$ 
as shown in the lower panels in Fig.~\ref{fig:Fig6}. 
This is a qualitative difference between the $s_{\pm}$-wave and $s_{++}$-wave pairing. 

The above difference originates from the difference of the mechanism of the enhancement between the two cases. 
In the $s_{\pm}$-wave pairing state, the enhancement originates from the sign change of the gap across the wave vector that bridges the Fermi surface, so that the maximum value of the ratio in $\Vec{Q}$-space is localized around this wave  vector. 
On the other hand, in the $s_{++}$-wave pairing state, the enhancement comes from the quasiparticle damping effect, which is not momentum dependent. 
As pointed out in previous studies\cite{Onari2,NagaiKuroki}, the dissipation increases in the high energy region ($E > 3 \Delta_{0}$) is the origin of the decrease of the intensity in the high energy region, so that 
one can see the hump structure as shown in Fig.~\ref{fig:Fig5} in the mid-range energy region ($2 \Delta_{0} < E < 3 \Delta_{0}$). 
This means that the enhancement (the superconducting to normal ratio) 
in the $s_{++}$-wave state and the nesting vector are not related. 
This was the original motivation for the proposal to investigate experimentally the wave vector $\sim (\pi,\pi)$ to distinguish between two 
cases\cite{NagaiKuroki}. 
In the present study, it can be more clearly  seen that the ratio of the enhancement in the $s_{++}$-wave state depends weakly on the wave vector. 
In addition, we also point out that this ratio does not 
depend on the doping level. 
This result is easily understood by the nature of the quasiparticle damping. 
Thus, if the experimentally observed peak structure is coming from the 
quasiparticle damping, 
the enhancement ratio of the intensity in the superconducting state 
to that in normal state has to be weakly momentum dependent.
This conclusion is valid  in various kinds of the iron-based superconductors.

\section{Conclusion}
In conclusion,  we have obtained the $\Vec{Q}$-map of the spin susceptibility in the $s_{\pm}$ and $s_{++}$ superconducting states of the iron-based superconductors by applying multiorbital RPA to the effective five-orbital models and considering the quasiparticle damping. 
We have shown that the peak position of the intensity shifts from the position on the line $Q_{x} = \pi$ to that 
on the line $Q_{y} = 0$ with decreasing the doping level. 
Considering the difficulty in quantitatively comparing the theories with the experiments, we have proposed that the comparison of the $\Vec{Q}$-dependence of the superconducting to normal ratio of the spin susceptibility is useful in 
determining the pairing state. 
In the $s_{\pm}$-wave state, the ratio is peaked around the nesting vector, 
and it is dependent on the doping level, while  
in the $s_{++}$-wave pairing state, the ratio is weakly dependent on the wave vector. Exploiting the present proposal, we expect that one can distinguish the 
two pairing states.

\section*{Acknowledgment }
We thank M. Machida, N. Nakai, Y. Ota, for helpful discussions and comments. 
The calculations have been performed using the supercomputing system PRIMERGY BX900 at the Japan Atomic Energy Agency.

\end{document}